\documentclass[]{pasj02}
\usepackage[switch,mathlines]{lineno} 
\usepackage{multirow}

\jyear{2025}
\Received{}
\Accepted{}

\begin{document} 

\title{Near-infrared [P~\emissiontype{II}] and [Fe~\emissiontype{II}] line mapping of Galactic supernova remnants}

\author{
 Takuma \textsc{Kokusho},\altaffilmark{1}\altemailmark \email{kokusho@u.phys.nagoya-u.ac.jp} 
 Yuki \textsc{Katsurada},\altaffilmark{1}
 Yong-Hyun \textsc{Lee},\altaffilmark{2}\orcid{0000-0003-3277-2147}
 Bon-Chul \textsc{Koo},\altaffilmark{3}\orcid{0000-0002-2755-1879}
 Takahiro \textsc{Nagayama},\altaffilmark{4}
 Hidehiro \textsc{Kaneda},\altaffilmark{1}\orcid{0000-0001-6879-1556}
 Koji S. \textsc{Kawabata},\altaffilmark{5,6}\orcid{0000-0001-6099-9539}
 Tatsuya \textsc{Nakaoka},\altaffilmark{5}
 Ho-Gyu \textsc{Lee},\altaffilmark{7}\orcid{0000-0002-3808-7143}
 and
 Rommy L.S.E. \textsc{Aliste Castillo}\altaffilmark{8}\orcid{0009-0009-2514-0062}
}
\altaffiltext{1}{Graduate School of Science, Nagoya University, Furo-cho, Chikusa-ku, Nagoya, Aichi 464-8602, Japan}
\altaffiltext{2}{Samsung SDS, Olympic-ro 35-gil 125, Seoul 05510, Republic of Korea}
\altaffiltext{3}{Department of Physics and Astronomy, Seoul National University, Seoul 08826, Republic of Korea}
\altaffiltext{4}{Graduate School of Science and Engineering, Kagoshima University, 1-21-35 Korimoto, Kagoshima 890-0065, Japan}
\altaffiltext{5}{Hiroshima Astrophysical Science Center, Hiroshima University, 1-3-1 Kagamiyama, Higashi-Hiroshima, Hiroshima 739-8526, Japan}
\altaffiltext{6}{Graduate School of Advanced Science and Engineering, Hiroshima University, 1-3-1 Kagamiyama, Higashi-Hiroshima, Hiroshima 739-8526, Japan}
\altaffiltext{7}{Korea Astronomy and Space Science Institute, Daejeon 34055, Republic of Korea}
\altaffiltext{8}{Planetary Atmospheres Group, Institute for Basic Science (IBS), Daejeon 34126, Republic of Korea}



\maketitle

\begin{abstract}
Phosphorus (P) is one of the key ingredients for life, yet its origins in galaxies remain poorly understood. In order to investigate the production of P by supernovae, we performed near-infrared (IR) [P~\emissiontype{II}] and [Fe~\emissiontype{II}] line mapping of $26$ Galactic supernova remnants (SNRs) with the Infrared Survey Facility and Kanata telescopes, using the narrow-band filters tuned to these lines. By combining our data with archival [Fe~\emissiontype{II}] maps from UKIRT, we detected both the [P~\emissiontype{II}] and [Fe~\emissiontype{II}] emissions in five SNRs, only the [Fe~\emissiontype{II}] emission in $15$ SNRs, and no line emissions in the remaining six. Using the observed [P~\emissiontype{II}]/[Fe~\emissiontype{II}] ratios and upper limits for non-detections, we derived the P/Fe abundance ratios, which vary by up to two orders of magnitude among our sample SNRs. This suggests that the production rate of P and/or the degree of  dust destruction may differ from remnant to remnant, the latter being due to the fact that P is volatile while Fe is mostly locked in dust grains. We used the mid- and far-IR maps to examine the dust content for the five SNRs where both the line emissions are detected. As a result, we find that high P/Fe abundance ratios in the northern and southeastern regions of Cassiopeia~A and the Crab Nebula, respectively, are not likely due to dust destruction but may reflect an asymmetric ejection of P during supernova explosions. In the Crab Nebula, it is also possible that near-IR [Ni~\emissiontype{II}] emission contaminates the observed flux in the southeastern region, suggesting that the Ni/Fe abundance ratio, rather than the P/Fe abundance ratio, is relatively high in this part of the remnant.
\end{abstract}


\section{Introduction} \label{sec:intro}

Phosphorus (P) is one of the key ingredients for life, playing an important role in biomolecules, such as deoxyribonucleic and ribonucleic acids. The mystery of P is that its abundance in the human body (the P/H number ratio${\sim}10^{-3}$; \cite{ems01}) is significantly higher than that in the interstellar medium (${\sim}10^{-7}$; \cite{asp09}), which may suggest an efficient mechanism for concentrating P from the interstellar space into biological systems. Most elements including P are produced inside stars and released into the interstellar space through stellar winds and/or supernovae. Stellar models predict that massive stars ($>10~M_{\solar}$) are more efficient in producing P than low-mass stars \citep{kob06, kar10}, suggesting that supernovae are likely to be the primary source of P in galaxies. P is thought to be mainly produced by explosive Ne-burning during the explosion of massive stars \citep{rau02, koo13}.

The evolution of P abundances in our galaxy can be traced through stars at different evolutionary stages. Previous studies measured P abundances in stars with a wide range of metallicities, to reveal that [P/Fe] decreases as [Fe/H] increases (e.g.,~\cite{caf11}). This result suggests that core-collapse supernovae are the primary source of P, because Fe is mainly produced by Type Ia supernovae, which occur at later evolutionary stages of galaxies (e.g.,~\cite{kob06}). However, the observed [P/Fe] is systematically higher than that expected from chemical evolution models of our galaxy over a wide range of [Fe/H] (e.g.,~\cite{ces12}), suggesting that these models may underestimate the production rate of P by supernovae. \citet{bek24} propose that P enrichment by novae can explain the observed trend between [P/Fe] and [P/H].

Freshly synthesized P in supernovae is found by \citet{koo13} who performed near-infrared (IR) spectroscopy of Cassiopeia~A (Cas~A), focusing on specific regions of its bright ejecta shell dominated by Ne- and O-burning elements. They measured P/Fe abundance ratios in Cas~A using the [P~\emissiontype{II}] $1.189$~$\micron$ and [Fe~\emissiontype{II}] $1.257$~$\micron$ emissions which have similar excitation temperatures, critical densities, and ionization potentials from neutral states, thereby facilitating abundance analyses. They find that the abundance ratio by number, $X$(P/Fe), in the bright ejecta shell of Cas~A is up to $100$ times higher than the solar abundance, in agreement with nucleosynthesis models. \citet{sol19} also detected the [P~\emissiontype{II}] $1.189$~$\micron$ emission from the Crab Nebula using spectroscopy, although they did not perform abundance analyses of P.

As such, observational evidence for the in-situ production of P by supernovae is still very limited, and the production rate of P in supernovae remains poorly understood. Previous studies of the [P~\emissiontype{II}] emission in SNRs rely on spectroscopy, and therefore, they may miss the [P~\emissiontype{II}] emission in some parts of the ejecta shell due to a limited field of view in spectroscopic observations. In addition, it is important to consider the depletion of Fe into dust when discussing $X$(P/Fe), because more than $99{\%}$ of Fe is locked in dust in the interstellar medium, while P is volatile (e.g.,~\cite{sav96}). Consequently, abundance analyses using $X$(P/Fe) in SNRs require careful consideration of dust destruction; observed $X$(P/Fe) can be lower than those in the general interstellar medium due to shocks destroying dust and releasing Fe into the gas phase. From this perspective, \citet{ali25} report a study of dust processing in SNRs in the Magellanic Clouds, using $X$(P/Fe) derived from the [P~\emissiontype{II}] and [Fe~\emissiontype{II}] emissions. These line emissions are also used to diagnose the ionizing mechanisms of the gas surrounding active galactic nuclei (AGN; \cite{oli01, sto09, cal23}).

In order to investigate the production of P by supernovae, we performed large-area mapping of the [P~\emissiontype{II}] and [Fe~\emissiontype{II}] emissions for $26$ Galactic SNRs, using the narrow-band filters tuned to these lines. From our data, we derived $X$(P/Fe) in radiative shock regions of our sample SNRs. In addition, we examine the degree of dust destruction by shocks, using the mid- and far-IR maps to better understand the origins of P in our sample SNRs. Section~\ref{sec:obs} describes details of our observation and data reduction, while section~\ref{sec:res} presents our observational results. Section~\ref{sec:dis} discusses the origins of P in each SNR, and section~\ref{sec:con} presents a summary of our study.

\section{Observation and data analysis} \label{sec:obs}

\begin{table*}[t]
\tbl{Specifications of the narrow-band filters used in this study}{
\begin{tabular}{lccc}
\hline
Transition & ${\lambda}_\mathrm{center}$ (\micron) & ${\Delta}{\lambda}_\mathrm{eff}$ (\micron)\footnotemark[$*$] & ${\Delta}v$ (km~s$^{-1}$)\footnotemark[$\dagger$] \\ 
\hline
[P~\emissiontype{II}]~$^1D_2~{\rightarrow}~^3P_2$ & $1.189$ & $0.028$ & $3500$ \\
$[\mathrm{Fe}\emissiontype{II}]~a^4D_{7/2}~{\rightarrow}~a^6D_{9/2}$ & $1.257$ & $0.028$ & $3300$ \\
$[\mathrm{Fe}\emissiontype{II}]~a^4D_{7/2}~{\rightarrow}~a^4F_{9/2}$ & $1.644$ & $0.031$ & $2800$ \\
Continuum for [P~\emissiontype{II}] and [Fe~\emissiontype{II}]$_{1.26}$ & $1.222$ & $0.026$ & -- \\
Continuum for [Fe~\emissiontype{II}]$_{1.64}$ & $1.570$ & $0.026$ & -- \\
\hline
\end{tabular}} \label{tab:fil}
\begin{tabnote}
\footnotemark[$*$] The effective band width is calculated by using ${\int}S(\lambda)d{\lambda}=T_{\lambda}{\Delta}{\lambda}_\mathrm{eff}$, where $S(\lambda)$ and $T_{\lambda}$ represent the filter response curve and throughput at ${\lambda}_\mathrm{center}$, respectively. \\
\footnotemark[$\dagger$] The velocity coverage corresponding to ${\Delta}{\lambda}_\mathrm{eff}$.
\end{tabnote}
\end{table*}

\begin{table*}[t]
\tbl{Summary of SNRs observed in our study}{
\begin{tabular}{ccccccccc}
\hline
Name & RA & Dec & \multicolumn{4}{c}{Exposure (min)} & Telescope & Date \\
& (J2000) & (J2000) & [P~\emissiontype{II}]\footnotemark[$*$] & [Fe~\emissiontype{II}]$_{1.26}$ & [Fe~\emissiontype{II}]$_{1.64}$ & Cont.\footnotemark[$\dagger$] & & \\
\hline
G1.9$+$0.3 & \timeform{17h48m45.4s} & \timeform{-27D10'06.2''} & 15 & $-$ & 15 & 15 & IRSF & 2016 May, 2023 Sep\\
Kepler & \timeform{17h30m40.9s} & \timeform{-21D29'23.9''} & 15 & $-$ & 15 & 15 & IRSF & 2016 May, 2023 Aug\\
G11.2--0.3 & \timeform{18h11m29.8s} & \timeform{-19D25'22.0''} & 100 & $-$ & L19\footnotemark[$\ddagger$] & 15 & IRSF & 2016 May, 2023 Sep\\
G15.9$+$0.2 & \timeform{18h18m53.4s} & \timeform{-15D01'27.0''} & 15 & $-$ & 15 & 15 & IRSF & 2016 May, 2023 Sep\\
Kes~73 & \timeform{18h41m19.9s} & \timeform{-04D56'15.1''} & 100 & $-$ & L19\footnotemark[$\ddagger$] & 15 & IRSF & 2016 May, 2023 Sep\\
Kes~75 & \timeform{18h46m25.6s} & \timeform{-02D58'38.0''} & 90 & $-$ & 90 & 15 & IRSF & 2016 May, 2023 Sep\\
3C~391 & \timeform{18h49m25.8s} & \timeform{00D56'11.2''} & 15 & $-$ & L19\footnotemark[$\ddagger$] & 15 & IRSF & 2016 May, 2023 Sep\\
W44 & \timeform{18h55m23.1s} & \timeform{01D20'00.1''} & 20 & $-$ & L19\footnotemark[$\ddagger$] & $-$ & IRSF & 2016 May\\
3C~396 & \timeform{19h04m06.3s} & \timeform{05D27'05.0''} & 15 & $-$ & L19\footnotemark[$\ddagger$] & 15 & IRSF & 2016 May, 2023 Sep\\
3C~397 & \timeform{19h07m35.1s} & \timeform{07D08'42.6''} & 15 & $-$ & L19\footnotemark[$\ddagger$] & 15 & IRSF & 2016 May, 2023 Sep\\
G41.5$+$0.4 & \timeform{19h05m38.1s} & \timeform{07D46'58.7''} & 15 & $-$ & L19\footnotemark[$\ddagger$] & 15 & IRSF & 2016 May, 2023 Sep\\
W49B & \timeform{19h11m07.3s} & \timeform{09D06'12.8''} & 15 & $-$ & L19\footnotemark[$\ddagger$] & 15 & IRSF & 2016 May, 2023 Sep\\
Cas~A & \timeform{23h23m25.3s} & \timeform{58D48'59.7''} & 80 & 120 & L19\footnotemark[$\ddagger$] & 120 & Kanata & 2021 Oct, 2022 Dec, 2023 Nov \\
Tycho & \timeform{00h25m15.7s} & \timeform{64D08'15.2''} & 100 & $-$ & 100 & $-$ & Kanata & 2021 Oct\\
Crab & \timeform{05h34m32.3s} & \timeform{22D01'11.3''} & 120 & 70 & 120 & 120 & IRSF & 2023 Feb, Aug, Sep\\
IC~443 (a) & \timeform{06h17m39.1s} & \timeform{22D50'18.3''} & 20 & K13\footnotemark[$\ddagger$] & $-$ & 20 & Kanata & 2021 Oct, 2022 Nov\\
IC~443 (b) & \timeform{06h17m37.3s} & \timeform{22D22'33.8''} & 20 & K13\footnotemark[$\ddagger$] & $-$ & 20 & Kanata & 2021 Oct, 2022 Nov\\
Puppis~A & \timeform{08h21m26.6s} & \timeform{-42D42'15.4''} & 120 & 30 & 120 & 120 & IRSF & 2016 May, 2023 Feb\\
G290.1--0.8 & \timeform{11h02m59.0s} & \timeform{-60D53'44.7''} & 15 & $-$ & 15 & $-$ & IRSF & 2016 May\\
G292.0$+$1.8 & \timeform{11h24m29.4s} & \timeform{-59D15'36.6''} & 70 & $-$ & 70 & $-$ & IRSF & 2016 May\\
Kes~17 & \timeform{13h06m01.1s} & \timeform{-62D40'53.2''} & 15 & $-$ & 15 & 15 & IRSF & 2016 May, 2023 Sep\\
RCW~86 (a) & \timeform{14h40m27.2s} & \timeform{-62D39'05.0''} & 15 & 7.5 & 15 & 15 & IRSF & 2016 May, 2023 Sep\\
RCW~86 (b) & \timeform{14h42m09.0s} & \timeform{-62D11'27.0''} & 15 & 30 & 15 & 15 & IRSF & 2014 Mar, 2016 May, 2023 Sep\\
MSH~15--52 & \timeform{15h13m18.6s} & \timeform{-59D02'27.5''} & 60 & $-$ & 60 & 60 & IRSF & 2016 May, 2023 Sep\\
RCW~103 & \timeform{16h17m30.5s} & \timeform{-51D03'17.0''} & 60 & 20 & 60 & 53 & IRSF & 2016 May, 2023 Sep\\
G337.2--0.7 & \timeform{16h39m31.0s} & \timeform{-47D49'40.3''} & 15 & $-$ & 15 & 15 & IRSF & 2016 May, 2023 Aug\\
G344.7--0.1 & \timeform{17h03m52.4s} & \timeform{-41D42'53.1''} & 15 & $-$ & 15 & 15 & IRSF & 2016 May, 2023 Aug\\
G349.7$+$0.2 & \timeform{17h17m59.6s} & \timeform{-37D26'20.8''} & 15 & $-$ & 15 & 15 & IRSF & 2016 May, 2023 Aug\\
\hline
\end{tabular}} \label{tab:obj}
\begin{tabnote}
\footnotemark[$*$] Typical $3{\sigma}$ detection limits of the [P~\emissiontype{II}] maps are $6{\times}10^{-6}$ and $3{\times}10^{-6}$~erg~s$^{-1}$~cm$^{-2}$~sr$^{-1}$ for exposure times shorter and longer than 30 minutes, respectively.\\
\footnotemark[$\dagger$] The [P~\emissiontype{II}] and [Fe~\emissiontype{II}]$_{1.64}$ continua were observed with the same exposure time. For the Crab Nebula, an additional observation with a $50$-min exposure was performed for the [Fe~\emissiontype{II}]$_{1.26}$ continuum.\\
\footnotemark[$\ddagger$] K13 and L19 indicate that the narrow-band images from \citet{kok13} and \citet{lee19} are used, respectively.\\ 
\end{tabnote}
\end{table*}

We performed [P~\emissiontype{II}] and [Fe~\emissiontype{II}] line mapping of Galactic SNRs using the Simultaneous Infrared Imager for Unbiased Survey (SIRIUS; \cite{nag99, nag03}) installed on the InfraRed Survey Facility (IRSF) $1.4$~m telescope, located at the South African Astronomical Observatory, and the Hiroshima Optical and Near-Infrared camera (HONIR; \cite{aki14}) installed on the $1.5$~m Kanata telescope, located at the Higashi Hiroshima Observatory. The SIRIUS and HONIR cameras have field of views of $\timeform{7.7'} \times \timeform{7.7'}$ and $\timeform{10'} \times \timeform{10'}$ with pixel scales of \timeform{0.45"} and \timeform{0.295"}, respectively. For line mapping observations, we used narrow-band filters tuned to the [P~\emissiontype{II}] $1.189~\micron$, [Fe~\emissiontype{II}] $1.257~\micron$, and [Fe~\emissiontype{II}] $1.644~\micron$ lines (hereafter referred to as  [P~\emissiontype{II}], [Fe~\emissiontype{II}]$_{1.26}$, and [Fe~\emissiontype{II}]$_{1.64}$, respectively) with the velocity coverage of  typically ${\pm}3000$~km~s$^{-1}$. We also used narrow-band filters designed to observe continuum emission at wavelengths adjacent to these target lines. Table~\ref{tab:fil} summarizes specifications of these filters. Possible contamination from other lines in our narrow-band filters is described in \citet{ali25}. According to their estimates, such contamination is negligible and/or not strong enough to affect our conclusions. However, in the Crab Nebula, the [Ni~\emissiontype{II}] $1.191~\micron$ emission is detected significantly (e.g.,~\cite{hud90}), and this line may contaminate the [P~\emissiontype{II}] narrow-band images of the Crab Nebula, as well as those of other sample SNRs. This issue is discussed further in section~\ref{sec:ori}.

We observed $26$ Galactic SNRs listed in table~\ref{tab:obj}. For SNRs larger than the field of views of the cameras, we observed selected bright regions. Given observational constraints, such as sky conditions, we prioritized observing the [P~\emissiontype{II}], [Fe~\emissiontype{II}]$_{1.64}$, and their corresponding continuum emissions, since [Fe~\emissiontype{II}]$_{1.64}$ is less affected by interstellar dust extinction than [Fe~\emissiontype{II}]$_{1.26}$. Therefore, some of our sample SNRs lack the [Fe~\emissiontype{II}]$_{1.26}$ and/or continuum observations, although this does not affect our conclusion. Because the brightness of the Crab Nebula is known to vary over time due to its strong synchrotron emission (e.g.,~\cite{del19}), we observed its line and continuum emissions during the same observing run. We also considered the spectral index of the synchrotron emission of $-0.48$ for the Crab Nebula \citep{tem24} when subtracting the continuum emission. For SNRs covered in the UKIRT Widefield Infrared Survey for Fe$^+$ \citep{lee14, lee19}, we used the [Fe~\emissiontype{II}]$_{1.64}$ maps from this survey, as they provide higher sensitivity than our observations.

We adopted standard procedures to reduce our data as described in \citet{kok13}. In addition, we applied spatial high-pass filtering with a cut-off spatial frequency of \timeform{1'}, following the method described in \citet{kok20}. We performed photometric calibration of the narrow-band images using the Two Micron All Sky Survey (2MASS) Point Source Catalog \citep{skr06} and assuming that the magnitudes of stars in the narrow-band images correspond to those in the 2MASS broad-band images. The uncertainty of the calibration coefficient is typically ${\leq}3{\%}$ for our narrow-band images. To mitigate contamination from residuals of stars in the continuum-subtracted images, we performed point spread function (PSF) photometry and subtracted stars, following the method adopted by \citet{lee14}, where the STARFINDER code (\cite{dio20}) is used to construct a PSF model for each frame. Finally, we smoothed the continuum-subtracted images using a Gaussian kernel with a sigma of \timeform{2.3"}.

\section{Results} \label{sec:res}

\subsection{[P~\emissiontype{II}] and [Fe~\emissiontype{II}] maps} \label{sec:map}

\begin{figure*}[t]
\begin{center}
\includegraphics[width=\textwidth]{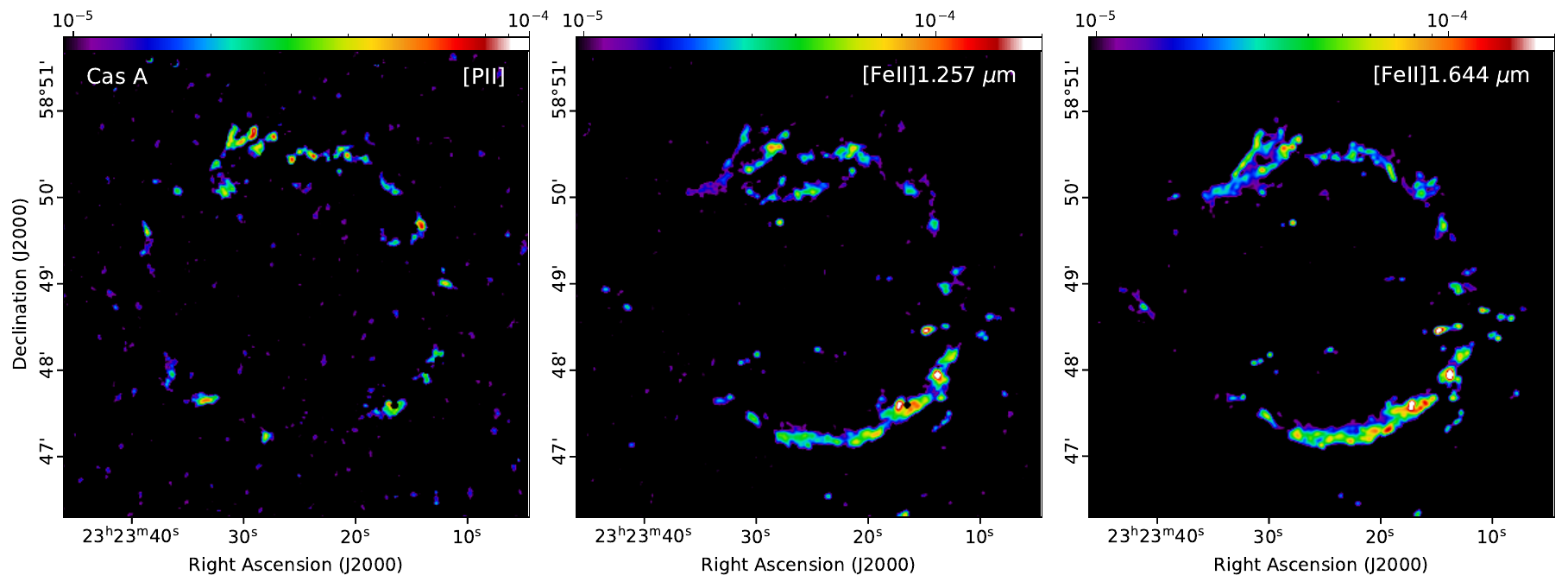}
\includegraphics[width=\textwidth]{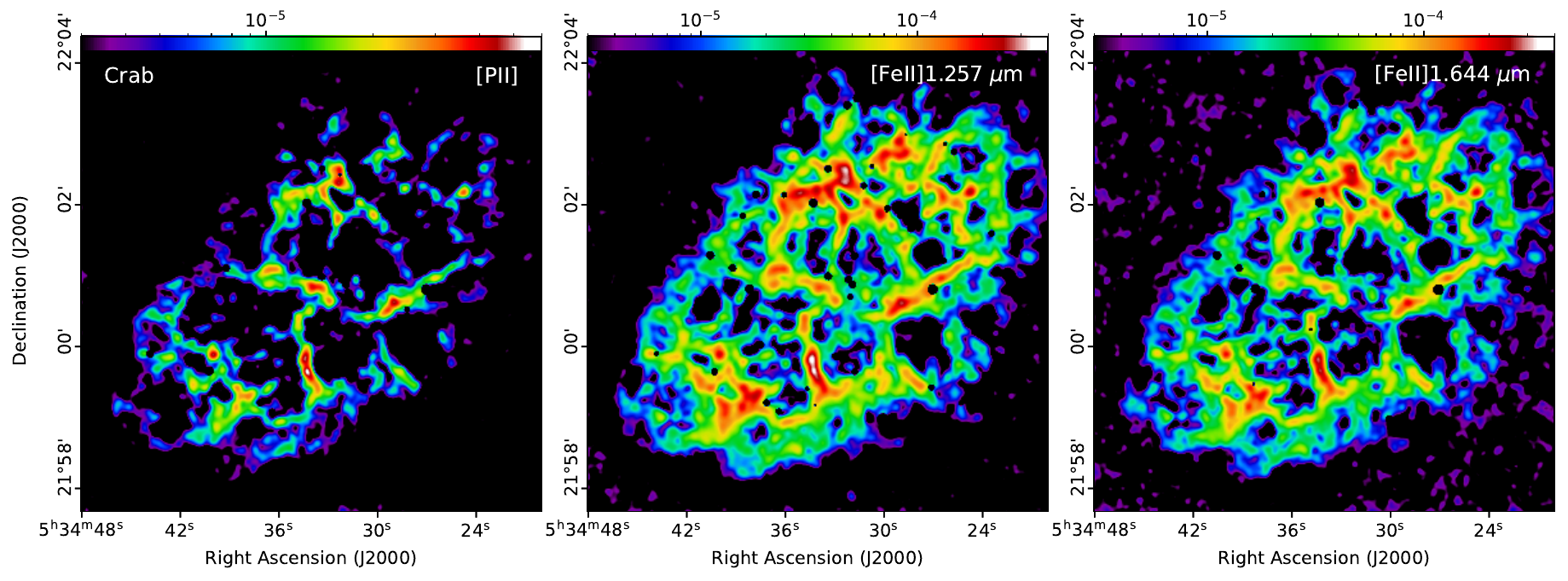}
\includegraphics[width=\textwidth]{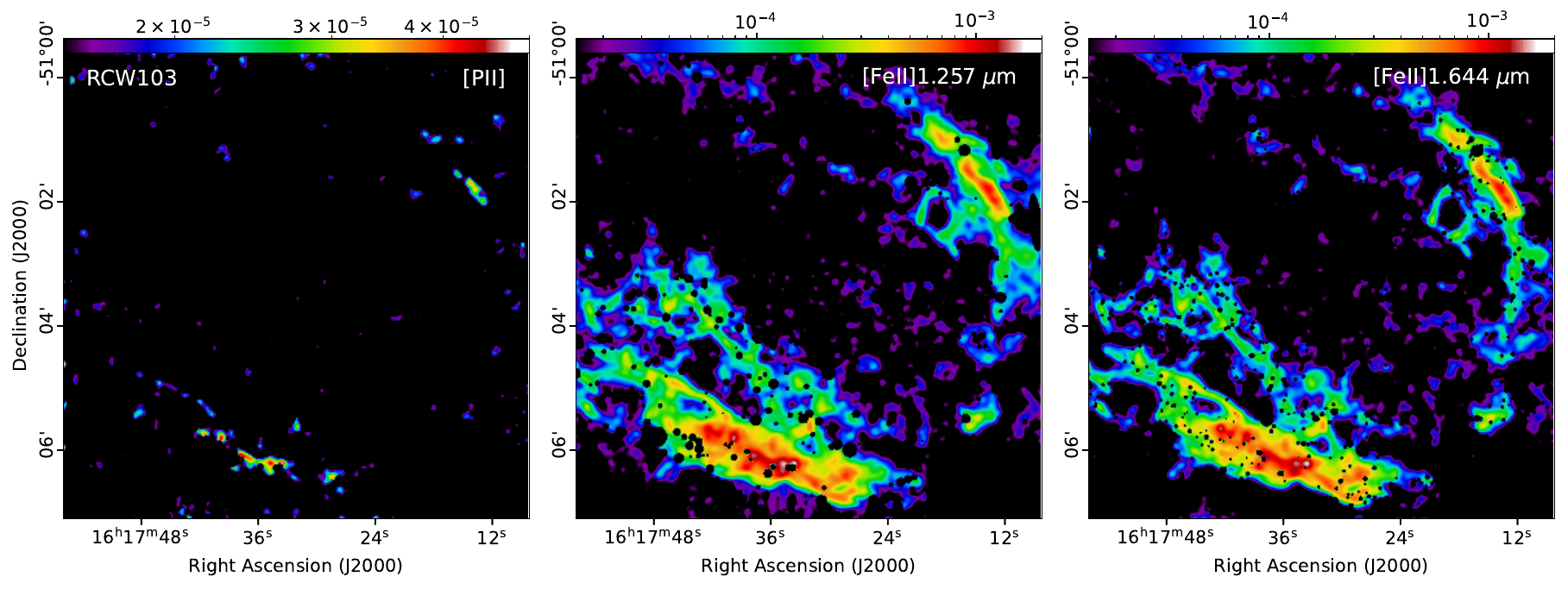}
\end{center}
\caption{Continuum-subtracted [P~\emissiontype{II}], [Fe~\emissiontype{II}]$_{1.26}$, and [Fe~\emissiontype{II}]$_{1.64}$ maps of Cas~A (top), the Crab Nebula (middle), and RCW~103 (bottom). The [Fe~\emissiontype{II}]$_{1.64}$ map of Cas~A is taken from \citet{koo18}. The color levels are in units of erg~s$^{-1}$~cm$^{-2}$~sr$^{-1}$. {Alt text: Nine maps.}} \label{fig:map1}
\end{figure*}

\begin{figure*}[t]
\begin{center}
\includegraphics[width=\textwidth]{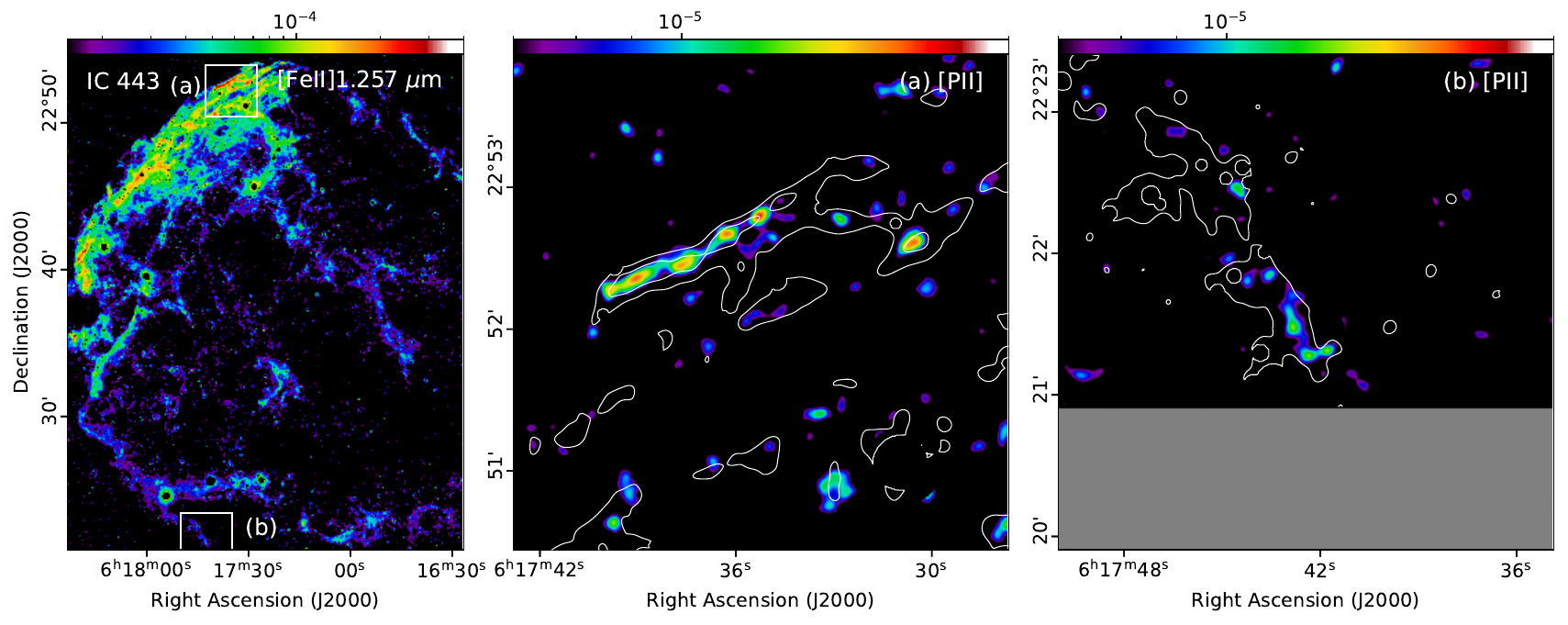}
\includegraphics[width=\textwidth]{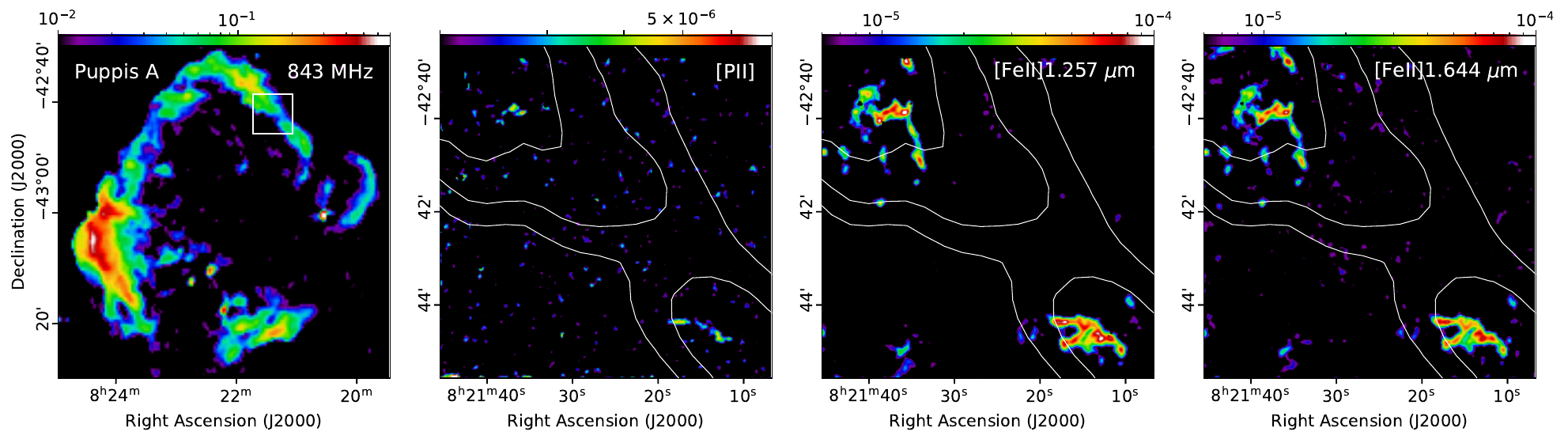}
\end{center}
\caption{Top: continuum-subtracted [P~\emissiontype{II}] and [Fe~\emissiontype{II}]$_{1.26}$ maps of IC~443. The [Fe~\emissiontype{II}]$_{1.26}$ map is taken from \citet{kok13}, where white boxes represent the target regions in our study. White contours in the [P~\emissiontype{II}] maps represent the [Fe~\emissiontype{II}]$_{1.26}$ emission. Bottom: $843$~GHz and continuum-subtracted [P~\emissiontype{II}], [Fe~\emissiontype{II}]$_{1.26}$, and [Fe~\emissiontype{II}]$_{1.64}$ maps of Puppis~A. The $843$~GHz map is taken from the SUMSS survey (\cite{mau03}), where a white box represents the target region in our study. White contours in the [P~\emissiontype{II}] and [Fe~\emissiontype{II}] maps represent the $843$~GHz emission. The color levels are in units of Jy/beam and erg~s$^{-1}$~cm$^{-2}$~sr$^{-1}$ for the radio and near-IR line maps, respectively. {Alt text: Seven maps.}} \label{fig:map2}
\end{figure*}

\begin{figure*}[t]
\begin{center}
\includegraphics[width=\textwidth]{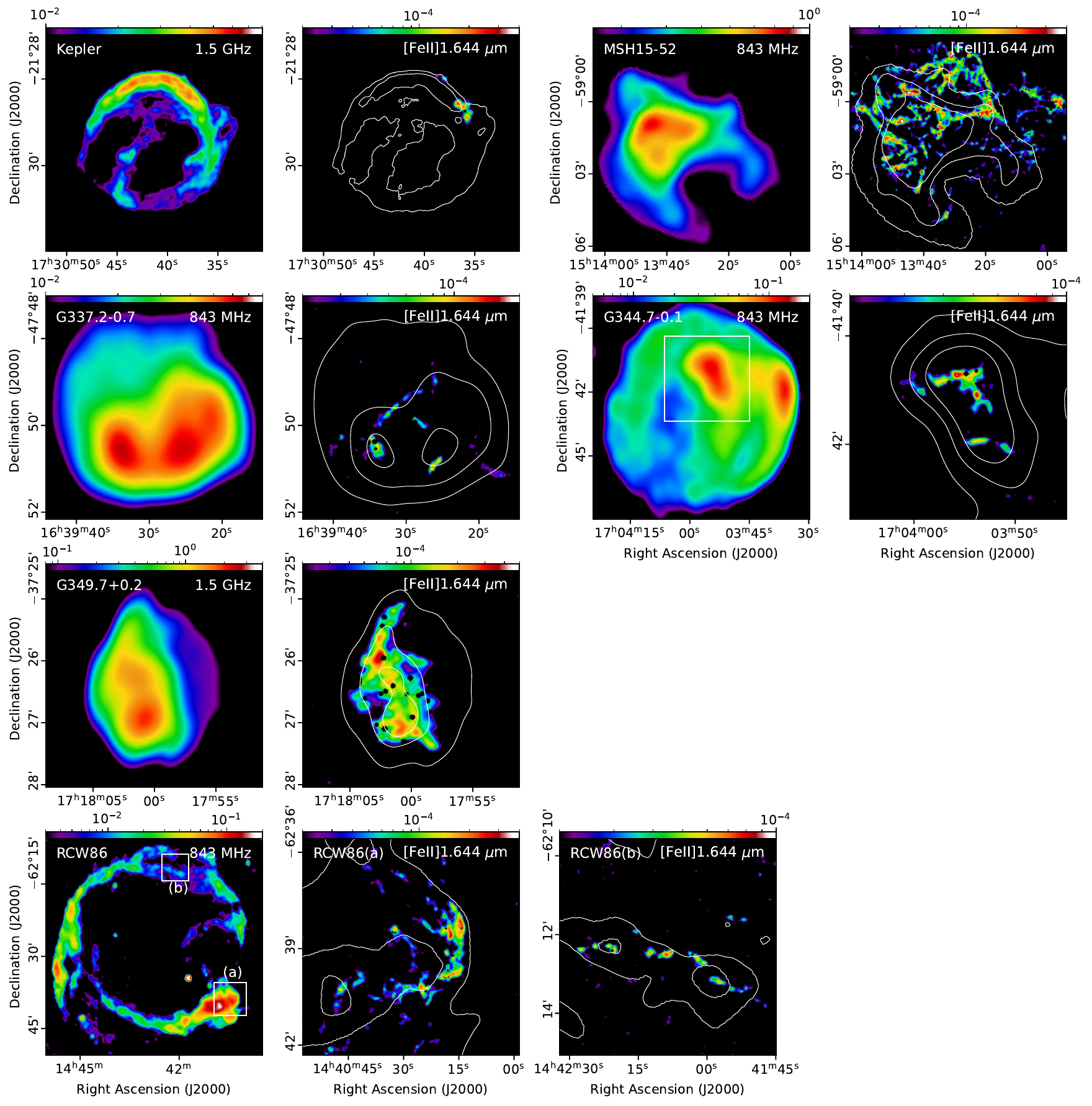}
\end{center}
\caption{Continuum-subtracted [Fe~\emissiontype{II}]$_{1.64}$ and radio continuum maps of Kepler, MSH~15--52, G337.2--0.7, G344.7--0.1, G349.7$+$0.2, and RCW~86. The radio continuum maps at $843$~MHz and $1.5$~GHz are taken from the SUMSS survey \citep{mau03} and NRAO VLA archive survey at $\langle$http://www.vla.nrao.edu/astro/nvas$\rangle$, respectively. White boxes in the radio maps of G344.7--0.1 and RCW~86 show the target region in our study. White contours in the [Fe~\emissiontype{II}]$_{1.64}$ maps represent the radio continuum emission. The color levels are in units of Jy/beam and erg~s$^{-1}$~cm$^{-2}$~sr$^{-1}$ for the radio and [Fe~\emissiontype{II}]$_{1.64}$ maps, respectively. {Alt text: 13 maps.}} \label{fig:map3}
\end{figure*}

\begin{figure}[t]
\begin{center}
\includegraphics[width=0.5\textwidth]{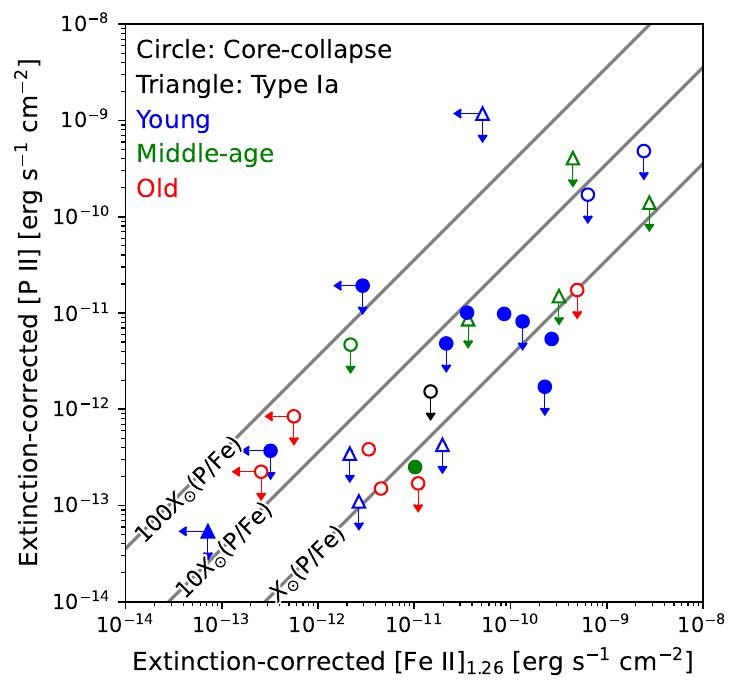}
\end{center}
\caption{Scatter plot between the extinction-corrected [P~\emissiontype{II}] and [Fe~\emissiontype{II}] fluxes. Circle and triangle represent core-collapse and Type Ia SNRs, respectively. Blue, green, and red represent the SNR ages of $<5000$ (young), $5000$--$10000$ (middle-age), and $>10000$~yr (old), respectively. The black data point corresponds to G41.5$+$0.4, for which the age is not estimated in previous studies. Filled and open symbols indicate exposure times longer and shorter than 30 minutes, respectively. Arrows represent $3{\sigma}$ upper limits. Gray lines represent the relation expected from corresponding $X$(P/Fe) (see text for details). {Alt text: A scatter plot with three lines.}} \label{fig:flux}
\end{figure}

Among our sample SNRs, we detected both the [P~\emissiontype{II}] and [Fe~\emissiontype{II}] emissions in five SNRs as shown in figures~\ref{fig:map1} and \ref{fig:map2}. While Cas~A shows bright [P~\emissiontype{II}] emission as previously identified by spectroscopy \citep{koo13}, we newly find that the [P~\emissiontype{II}] emission is distributed across the entire shock region in Cas~A. The [Fe~\emissiontype{II}]$_{1.26}$ and [Fe~\emissiontype{II}]$_{1.64}$ maps of Cas~A are not identical, particularly in the northern region where the [Fe~\emissiontype{II}]$_{1.26}$ emission is brighter than the [Fe~\emissiontype{II}]$_{1.64}$ emission. This discrepancy is likely due to the different velocity coverage of the filters: our [Fe~\emissiontype{II}]$_{1.26}$ map covers velocities up to ${\pm}3300$~km~s$^{-1}$, while the [Fe~\emissiontype{II}]$_{1.64}$ map taken by \citet{koo18} only covers up to ${\pm}2600$~km~s$^{-1}$. The line-of-sight velocity in the northern region in Cas~A is ${\lesssim}5000$~km~s$^{-1}$ (e.g.,~\cite{mil15}), and thus the [Fe~\emissiontype{II}]$_{1.64}$ map misses part of the line emission in this region. On the other hand, the [P~\emissiontype{II}] and [Fe~\emissiontype{II}]$_{1.26}$ maps have comparable velocity coverage (see table~\ref{tab:fil}). Therefore, this effect is likely negligible in our P/Fe abundance analysis using the [P~\emissiontype{II}] and [Fe~\emissiontype{II}]$_{1.26}$ maps presented in section~\ref{sec:abu}.

For the other four SNRs (the Crab Nebula, RCW~103, IC~443, and Puppis~A) in figures~\ref{fig:map1} and \ref{fig:map2}, we detected the [P~\emissiontype{II}] emission for the first time. In general, the [P~\emissiontype{II}] and [Fe~\emissiontype{II}] emissions are co-spatial, and the [P~\emissiontype{II}] emission is significantly fainter than the [Fe~\emissiontype{II}] emission. On the other hand, the [P~\emissiontype{II}] emission is as strong as the  [Fe~\emissiontype{II}] emission in Cas~A, and the spatial distributions of the two line emissions slightly differ at small spatial scales, which will be discussed later. For a consistency check, we compared the observed [P~\emissiontype{II}]/[Fe~\emissiontype{II}] flux ratio derived from the narrow-band images with that obtained from spectroscopy at the eight slit positions of Cas~A \citep{koo13}. The ratios are in good agreement with a typical scatter of $30{\%}$, likely due to uncertainties in the measured fluxes and the different spatial resolutions of the observations. For the bright shell of RCW~103, \citet{oli90} derived an upper limit on the [P~\emissiontype{II}]/[Fe~\emissiontype{II}] flux ratio from spectroscopy, and we find that this limit is consistent with the ratio measured from the narrow-band images. We also newly detected the [Fe~\emissiontype{II}] emission in six SNRs (Kepler, MSH~15--52, G337.2--0.7, G344.7--0.1, G349.7$+$0.2, and RCW~86) shown in figure~\ref{fig:map3}, which were not covered in the UKIRT [Fe~\emissiontype{II}] survey \citep{lee14}. By combining our results with the UKIRT [Fe~\emissiontype{II}] survey, we detected only the [Fe~\emissiontype{II}] emission in $15$ SNRs and no line emissions in the remaining six.

We performed aperture photometry to measure the [P~\emissiontype{II}] and [Fe~\emissiontype{II}] fluxes for each SNR, using a circular or an elliptical aperture encompassing the line-emitting regions. For SNRs where no line emission was detected significantly, we used an aperture covering the radio continuum emission to obtain the upper limits of the line fluxes. We corrected the measured line fluxes for dust extinction as described below. As the [Fe~\emissiontype{II}]$_{1.26}$ and [Fe~\emissiontype{II}]$_{1.64}$ lines share the same upper level, their intrinsic [Fe~\emissiontype{II}]$_{1.26}$/[Fe~\emissiontype{II}]$_{1.64}$ flux ratio is fixed at $1.36$ (e.g.,~\cite{deb11}). Comparing this intrinsic ratio to the observed one and assuming the interstellar dust extinction law of \citet{car89}, we can estimate dust extinction toward SNRs detected in both the [Fe~\emissiontype{II}]$_{1.26}$ and [Fe~\emissiontype{II}]$_{1.64}$ emissions. To estimate dust extinction, we adopted the same method as \citet{kok13} who derived the dust extinction map from the [Fe~\emissiontype{II}]$_{1.26}$/[Fe~\emissiontype{II}]$_{1.64}$ flux ratio for IC~443. This method cannot be applied to Cas~A due to the different velocity coverage of the [Fe~\emissiontype{II}]$_{1.26}$ and [Fe~\emissiontype{II}]$_{1.64}$ maps as mentioned above. Instead, we used the same method as \citet{koo18} who derived the conversion factor between the total gas column density ($N_\mathrm{H}$) and the $250~\micron$ flux density for Cas~A. By applying this conversion to the Herschel $250~\micron$ map, we derived the $N_\mathrm{H}$ map of Cas~A. We then converted $N_\mathrm{H}$ to $A_V$ using the relation of $N_\mathrm{H}/A_V=1.87{\times}10^{21}$~cm$^{-2}$~mag$^{-1}$ \citep{dra03}. We also applied this $N_\mathrm{H}/A_V$ relation to our sample SNRs not detected in both the [Fe~\emissiontype{II}]$_{1.26}$ and [Fe~\emissiontype{II}]$_{1.64}$ emissions and lacking available $A_V$ measurements in literatures, where $N_\mathrm{H}$ was derived from previous X-ray studies. Possible systematic errors in this method may lead to an overestimation of the extinction-corrected line fluxes by up to $60${\%} \citep{lee15, koo18}. The extinction-corrected line fluxes thus derived are summarized in table~\ref{tab:flux}.

Figure~\ref{fig:flux} shows a scatter plot between the extinction-corrected [P~\emissiontype{II}] and [Fe~\emissiontype{II}]$_{1.26}$ fluxes, where data points are classified according to the explosion types and SNR ages listed in table~\ref{tab:para} in the Appendix. For SNRs lacking the [Fe~\emissiontype{II}]$_{1.26}$ observation, we estimate their [Fe~\emissiontype{II}]$_{1.26}$ flux from the extinction-corrected [Fe~\emissiontype{II}]$_{1.64}$ flux, assuming an intrinsic [Fe~\emissiontype{II}]$_{1.26}$/[Fe~\emissiontype{II}]$_{1.64}$ flux ratio of $1.36$. The figure shows that the [P~\emissiontype{II}]/[Fe~\emissiontype{II}]$_{1.26}$ flux ratio varies by up to two orders of magnitude, indicating a significant difference in the P/Fe abundance ratio among our sample SNRs, while there is no clear trend depending on the explosion types or SNR ages. All five objects detected in the [P~\emissiontype{II}] emission are core-collapse SNRs, suggesting a possible relation between the explosion types and the presence of P. However, since the exposure times differ among our sample SNRs, it is difficult to systematically investigate possible relations between the detection rate of the line emission and the SNR properties. Indeed, the [P~\emissiontype{II}] detection rates are 4/10 and 1/16 for SNRs observed with exposure times longer and shorter than $30$~minutes, respectively, indicating that more uniform observations of the SNRs are needed to study such relations. Similarly, although the non-detection of the [Fe~\emissiontype{II}] emission suggests that dust destruction is not as severe as the [Fe~\emissiontype{II}]-detected SNRs and/or that Fe-rich ejecta are absent, more uniform observations of the SNRs are required to examine this possibility. The variation in the P/Fe abundance ratio is further examined in section~\ref{sec:abu}.

\begin{table}
\tbl{Extinction-corrected line fluxes of our sample SNRs.}{
\begin{tabular}{ccccc}
\hline
Name & $A_V$\footnotemark[$*$] & $F_\mathrm{[P~\emissiontype{II}]}$ & $F_\mathrm{[Fe~\emissiontype{II}]1.26}$\footnotemark[$\dagger$] & References\footnotemark[$\ddagger$] \\
& (mag) & \multicolumn{2}{c}{($10^{-12}$~erg~s$^{-1}$~cm$^{-2}$)} & \\
\hline
G1.9$+$0.3 & 33 & $<$1000 & $<$50 & (1)\\
Kepler & 2.5 & $<$0.1 & 2.67$\pm$0.03 & (2)\\
G11.2--0.3 & 14 & $<$8 & 133.8$\pm$0.1 & (3)\\
G15.9$+$0.2 & 18 & $<$5 & 2.18$\pm$0.03 & (4)\\
Kes~73 & 12 & $<$5 & 21.54$\pm$0.07 & (5)\\
Kes~75 & 18 & $<$20 & $<$3 & (6)\\
3C~391 & 13 & $<$20 & 494.5$\pm$0.2 & (7)\\
W44 & 0.54 & $<$0.2 & 11.00$\pm$0.02 & (8)\\
3C~396 & 22 & $<$200 & 637.4$\pm$0.7 & (9)\\
3C~397 & 14 & $<$10 & 319.6$\pm$0.2 & (10)\\
G41.5$+$0.4 & 7.4 & $<$2 & 14.81$\pm$0.05 & (11)\\
W49B & 23 & $<$100 & 2776.2$\pm$0.3 & (12)\\
Cas~A & 8.5 & 10.1$\pm$0.3 & 35.4$\pm$0.2 & This work\\
Tycho & 2.1 & $<$0.05 & $<$0.07 & (13)\\
Crab & 2.4 & 9.80$\pm$0.04 & 86.07$\pm$0.06 & This work\\
IC~443 (a) & 2.6 & 0.150$\pm$0.007 & 4.53$\pm$0.01 & (14)\\
IC~443 (b) & 3.0 & 0.38$\pm$0.03 & 3.37$\pm$0.03 & (14)\\
Puppis~A & 3.5 & 0.25$\pm$0.02 & 10.23$\pm$0.03 & This work\\
G290.1--0.8 & 3.3 & $<$0.8 & $<$0.6 & (15)\\
G292.0$+$1.8 & 2.7 & $<$0.4 & $<$0.3 & (16)\\
Kes~17 & 1.9 & $<$0.2 & $<$0.3 & (17)\\
RCW~86 (a) & 1.7 & $<$0.4 & 19.7$\pm$0.1 & This work\\
RCW~86 (b) & 1.5 & $<$0.3 & 2.1$\pm$0.1 & This work\\
MSH~15--52 & 5.4 & $<$2 & 227.8$\pm$0.4 & (18)\\
RCW~103 & 2.1 & 5.38$\pm$0.06 & 268.33$\pm$0.07 & This work\\
G337.2--0.7 & 14 & $<$9 & 36.3$\pm$0.7 & (19)\\
G344.7--0.1 & 29 & $<$400 & 446$\pm$5 & (20)\\
G349.7$+$0.2 & 29 & $<$500 & 2435$\pm$7 & (21)\\
\hline
\end{tabular}} \label{tab:flux}
\begin{tabnote}
\footnotemark[$*$] $A_V$ is estimated from $N_\mathrm{H}$ for SNRs with no available $A_V$ measurements. The median of $A_V$ maps is presented for SNRs detected in both the [Fe~\emissiontype{II}]$_{1.26}$ and [Fe~\emissiontype{II}]$_{1.64}$ emissions.\\
\footnotemark[$\dagger$] For SNRs not observed in the [Fe~\emissiontype{II}]$_{1.26}$ emission, the corresponding flux is estimated from the extinction-corrected [Fe~\emissiontype{II}]$_{1.64}$ flux by assuming an intrinsic $\mathrm{[Fe~\emissiontype{II}]}_{1.26}/\mathrm{[Fe~\emissiontype{II}]}_{1.64}$ flux ratio of $1.36$.\\
\footnotemark[$\ddagger$] References for $A_V$ or $N_\mathrm{H}$. ``This work'' indicates that the corresponding $A_V$ is estimated from the [Fe~\emissiontype{II}]$_{1.26}$/[Fe~\emissiontype{II}]$_{1.64}$ flux ratio.
(1) \citet{bor17a},
(2) \citet{kas21},
(3) \citet{bor16},
(4) \citet{rey06},
(5) \citet{kum14},
(6) \citet{hel03},
(7) \citet{sat14},
(8) \citet{she04},
(9) \citet{su11},
(10) \citet{saf05},
(11) \citet{lee19},
(12) \citet{hwa00},
(13) \citet{che80},
(14) \citet{kok13},
(15) \citet{kam15},
(16) \citet{bha19},
(17) \citet{was16},
(18) \citet{bor20},
(19) \citet{tak16},
(20) \citet{fuk20},
(21) \citet{yas14} \\
\end{tabnote}
\end{table}

\subsection{P/Fe abundance ratio} \label{sec:abu}

\begin{table*}[t]
\tbl{P/Fe abundance ratio of the SNRs detected in the [P~\emissiontype{II}] and [Fe~\emissiontype{II}] emissions.}{
\begin{tabular}{ccccc}
\hline
Name & $F_\mathrm{[P~\emissiontype{II}]}/F_\mathrm{[Fe~\emissiontype{II}]1.26}$ & $n_e$ & $X$(P/Fe) & References\footnotemark[$*$] \\
& & ($10^3$~cm$^{-3}$) & ($X_\odot$(P/Fe)) & \\
\hline
Cas~A & 0.285$\pm$0.008 & 1--100 & 5.3--15.1 & (1)\\
Crab & 0.1139$\pm$0.0005 & 5--15 & 3.4--4.4 & (2)\\
IC~443 (a) & 0.033$\pm$0.002 & 4--10 & 1.0--1.4 & (3)\\
IC~443 (b) & 0.114$\pm$0.009 & 4--10 & 3.5--5.0 & (3)\\
Puppis~A & 0.025$\pm$0.002 & 1--10 & 0.8--1.4 & (4)\\
RCW~103 & 0.0201$\pm$0.0002 & 4.5--6.7 & 0.7--0.8 & (5)\\
\hline
\end{tabular}} \label{tab:abu}
\begin{tabnote}
\footnotemark[$*$] References for $n_e$. (1) \citet{lee17}, (2) \citet{hud90, tem24}, (3) \citet{gra87, rho01, koo16}, (4) \citet{sut95}, (5) \citet{oli89, koo16} \\
\end{tabnote}
\end{table*}

\begin{figure}[t]
\begin{center}
\includegraphics[width=0.5\textwidth]{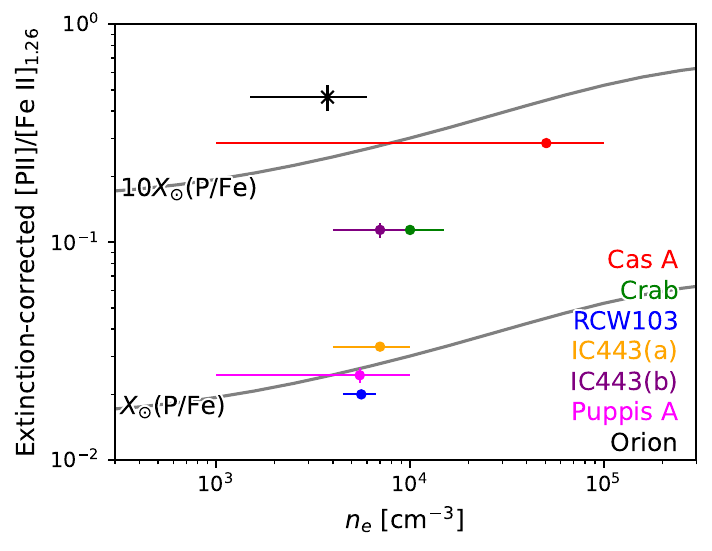}
\end{center}
\caption{Relations between the extinction-corrected [P~\emissiontype{II}]/[Fe~\emissiontype{II}]$_{1.26}$ flux ratio and $n_e$ of our sample SNRs detected in both the line emissions. Each SNR is represented by a circle in a different color, while the Orion bar is shown as a black cross for comparison \citep{wam00}. Gray lines represent relations expected from corresponding $X$(P/Fe). {Alt text: A scatter plot with two lines.}} \label{fig:ne}
\end{figure}

To investigate the P abundance of our sample SNRs, we estimate the P/Fe abundance ratio by number, $X$(P/Fe), using the method of \citet{koo13}. This method estimates $X$(P/Fe) from the [P~\emissiontype{II}]/[Fe~\emissiontype{II}]$_{1.26}$ flux ratio, utilizing the fact that the [P~\emissiontype{II}] and [Fe~\emissiontype{II}]$_{1.26}$ lines have similar excitation temperatures ($12780$~K for [P~\emissiontype{II}] and $11450$~K for [Fe~\emissiontype{II}]$_{1.26}$), critical densities for excitation by electron collisions ($5.3{\times}10^4$~cm$^{-3}$ for [P~\emissiontype{II}] and $3.5{\times}10^4$~cm$^{-3}$ for [Fe~\emissiontype{II}]$_{1.26}$), and ionization potentials from the neutral state ($10.5$~eV for P~\emissiontype{II} and $7.9$~eV for Fe~\emissiontype{II}). These similarities suggest that the two lines likely trace similar spatial regions, enabling more reliable abundance analysis. In addition, their line ratio is less affected by extinction owing to their similar wavelengths. 

According to \citet{koo13}, the [P~\emissiontype{II}]/[Fe~\emissiontype{II}]$_{1.26}$ flux ratio is expressed as
\begin{equation}
\frac{F_\mathrm{[P\emissiontype{II}]}}{F_\mathrm{[Fe\emissiontype{II}]1.26}} = a(n_e,~T_e)\left(\frac{f_\mathrm{P\emissiontype{II}}}{f_\mathrm{Fe\emissiontype{II}}}\right)X(\mathrm{P/Fe}),
\label{eq:rat1}
\end{equation}
where $a(n_e,~T_e)$ is a factor determined by the Einstein coefficients and the level population ratio ($f_{^1D_2, \mathrm{P\emissiontype{II}}}/f_{a^4D_{7/2}, \mathrm{Fe\emissiontype{II}}}$), while $n_e$ and $T_e$ are the electron density and temperature, respectively. Since the two lines have comparable excitation temperatures, $a(n_e,~T_e)$ mainly depends on $n_e$ \citep{koo13}. The parameters $f_\mathrm{P\emissiontype{II}}$ and $f_\mathrm{Fe\emissiontype{II}}$ are the fractional ionization of each element. In collisional ionization equilibrium, $f_\mathrm{P\emissiontype{II}}$ and $f_\mathrm{Fe\emissiontype{II}}$ peak near unity at $1.6{\times}10^4$ and $1.3{\times}10^4$~K, respectively. However, in shocked gas, the ionization fractions are affected by time-dependent collisional ionization and photoionization, and may significantly deviate from their equilibrium value. \citet{koo13} argued that, as the [P~\emissiontype{II}] and [Fe~\emissiontype{II}]$_{1.26}$ lines have similar physical properties, a simplified model assuming $T_e=10^4$~K, uniform density, and $f_\mathrm{P\emissiontype{II}}/f_\mathrm{Fe\emissiontype{II}}=1$ can yield $X$(P/Fe) to within a factor of two. We adopt this approach in our study. The dependence of $a(n_e,~T_e)$ on $n_e$ is discussed in \citet{koo13}, and for $n_e=10^2$--$10^6$~cm$^{-3}$, equation~(\ref{eq:rat1}) becomes
\begin{equation}
\frac{F_\mathrm{[P\emissiontype{II}]}}{F_\mathrm{[Fe\emissiontype{II}]1.26}} = (2\mathrm{-}7)X(\mathrm{P/Fe}).
\label{eq:rat2}
\end{equation}

In the Crab Nebula, the [P~\emissiontype{II}] and [Fe~\emissiontype{II}] emissions are thought to originate from the dense cores of filaments, which are ionized from the outside inward by synchrotron emission (e.g.,~\cite{hes08}). One approach to estimating $f_\mathrm{P\emissiontype{II}}$ and $f_\mathrm{Fe\emissiontype{II}}$ in the Crab Nebula is to adopt results from AGNs, where the gas is photoionized by a power-law continuum. \citet{cal23} used the photoionization code CLOUDY to model near-IR line emissions including [P~\emissiontype{II}] and [Fe~\emissiontype{II}] in both star-forming and AGN powered-galaxies. They derived the coefficient in equation (\ref{eq:rat2}) to be in the range of $3.4$--$5.4$ for hydrogen densities of $10^2$--$10^4$~cm$^{-3}$, which is comparable to the estimates by \citet{koo13} for shocked gas, with some discrepancies likely arising from differences in the adopted atomic parameters. \citet{hud90} estimated $n_e$ of the Crab Nebula using near-IR [Fe\emissiontype{II}] emissions, assuming $T_e=1.3{\times}10^4$~K. This estimate was updated by \citet{tem24} using revised atomic parameters and a model of the nebula radiation field, yielding $n_e=5{\times}10^3$--$1.5{\times}10^4$~cm$^{-3}$, which falls within the parameter range assumed by \citet{koo13} for shocked gas. Therefore, we conclude that equations (\ref{eq:rat1}) and (\ref{eq:rat2}) can reasonably be applied to the Crab Nebula for estimating $X$(P/Fe).

To obtain a rough estimate of $X$(P/Fe) from our photometry results, the gray lines in figure~\ref{fig:flux} show the relations expected from the solar P/Fe abundance of $X_{\odot}\mathrm{(P/Fe)}=8.1{\times}10^{-3}$ \citep{asp09}, $10X_{\odot}$(P/Fe), and $100X_{\odot}$(P/Fe), assuming a typical value of $a(n_e,~T_e)=4.5$ in equation~(\ref{eq:rat2}). To evaluate $X$(P/Fe) more precisely, we perform the abundance analysis for the five SNRs detected in both the [P~\emissiontype{II}] and [Fe~\emissiontype{II}]$_{1.26}$ emissions. We estimate their $a(n_e,~T_e)$ using $n_e$ from previous studies, which are primarily derived from near-IR [Fe~\emissiontype{II}] lines, and assuming $T_e=10^4$~K. Table~\ref{tab:abu} lists the $X$(P/Fe) thus derived for the five SNRs, and figure~\ref{fig:ne} shows their [P~\emissiontype{II}]/[Fe~\emissiontype{II}]$_{1.26}$ flux ratio as a function of $n_e$, along with the corresponding $X$(P/Fe). These results show that $X$(P/Fe) varies significantly among the five SNRs by more than an order of magnitude. For reference, we also include the Orion Bar in figure~\ref{fig:ne} \citep{wam00}, which shows the highest [P~\emissiontype{II}]/[Fe~\emissiontype{II}]$_{1.26}$ flux ratio in the plot, likely due to the depletion of Fe into dust. Hence, the scatter observed among the SNRs in figure~\ref{fig:ne} also suggests that the degree of dust destruction varies from remnant to remnant. This is likely the case especially for RCW~103, IC~443, and Puppis~A, which are known to be interacting with the surrounding interstellar/circumstellar medium (e.g.~\cite{are10, mil21}). This point will be further discussed in section~\ref{sec:ir}. Cas~A shows $X$(P/Fe) of ${\sim}10X_{\odot}$(P/Fe), which is lower than the estimates by \citet{koo13} for the P-rich knots in Cas~A, because our measurement is based on the photometry of the entire remnant and thus likely diluted by regions poor in P.

\subsection{Comparison with previous studies} \label{sec:com}

{
\setlength{\parskip}{0pt}

For the SNRs newly detected in the [Fe~\emissiontype{II}] emission in our study, we compare their [Fe~\emissiontype{II}] maps with previous results to better understand the origin of the [Fe~\emissiontype{II}] emission. A similar comparison for the SNRs detected in both the [P~\emissiontype{II}] and [Fe~\emissiontype{II}] emissions is presented later.

The [Fe~\emissiontype{II}]-bright shell in Kepler is also bright in the mid-IR emission, and \citet{bla07} suggest that circumstellar dust is interacting with fast shocks in this region. Indeed, spectroscopy of part of this circumstellar knot detected the [Fe~\emissiontype{II}] emission \citep{ger01}, supporting the dust destruction scenario. For part of the [Fe~\emissiontype{II}]-bright shell in MSH~15--52, \citet{sew83} detected the [Fe~\emissiontype{II}] emission using spectroscopy, and they discuss interactions between shocks and the interstellar medium. On the other hand, X-ray analyses of the SNR suggest that ejecta knots are interacting with the ambient medium around the northern part of the [Fe~\emissiontype{II}]-bright shell. In addition, the [Fe~\emissiontype{II}] shell is likely associated with the mid-IR source near the center of the SNR, which is possibly an ejecta knot mixed with circumstellar material \citep{koo11, kim24}. Therefore, the [Fe~\emissiontype{II}] emission in MSH~15--52 may originate from ejecta and/or ambient material. In G337.2--0.7, a bright ring structure is identified at radio wavelengths, which is likely of interstellar/circumstellar origin based on abundance analyses of X-ray spectra \citep{rak06}. The [Fe~\emissiontype{II}] shell found in our study is well correlated with the radio ring, suggesting that the [Fe~\emissiontype{II}] emission in this SNR originates from dust destruction. For the [Fe~\emissiontype{II}]-bright shell in G344.7--0.1, mid-IR imaging and spectroscopy suggest that shocks are interacting with the interstellar medium \citep{rea06, and11}. In particular, \citet{and11} analyze the mid-IR line ratios and conclude that fast shocks with velocities of ${\gtrsim}100$~km~s$^{-1}$ are present, suggesting that the [Fe~\emissiontype{II}] emission is of interstellar origin. In G349.7$+$0.2, molecular clouds interacting with shocks are identified through radio observations near the [Fe~\emissiontype{II}]-bright shell \citep{rey00, rey01, laz10}. Furthermore, mid-IR spectroscopy of the SNR suggests dust destruction, indicating that the [Fe~\emissiontype{II}] emission is of interstellar origin \citep{and11}. For the two regions detected in the [Fe~\emissiontype{II}] emission in RCW~86, \citet{wil11} investigate the mid-IR and X-ray emissions and conclude that interstellar dust is heated by shocks in these regions, suggesting that the [Fe~\emissiontype{II}] emission is of interstellar origin. On the other hand, X-ray analyses of the shells also reveal the presence of low-ionized Fe of ejecta origin, possibly suggesting that part of the [Fe~\emissiontype{II}] emission may arise from supernova ejecta.

Overall, most of the [Fe~\emissiontype{II}] emissions detected in these SNRs are likely associated with interstellar/circumstellar components interacting with shocks, suggesting that the [Fe~\emissiontype{II}] emission originates from dust destruction by shocks. However, for MSH~15--52 and RCW~86, there remains a possibility that part of the emission arises from Fe-rich ejecta. Since ejecta components tend to show higher velocities than interstellar components, future spectroscopic observations are required to discriminate between these possibilities.

}



\section{Discussion} \label{sec:dis}

\subsection{Effects of dust destruction on the [P~\emissiontype{II}]/[Fe~\emissiontype{II}] ratio} \label{sec:ir}

\begin{figure*}[t]
\begin{center}
\includegraphics[width=\textwidth]{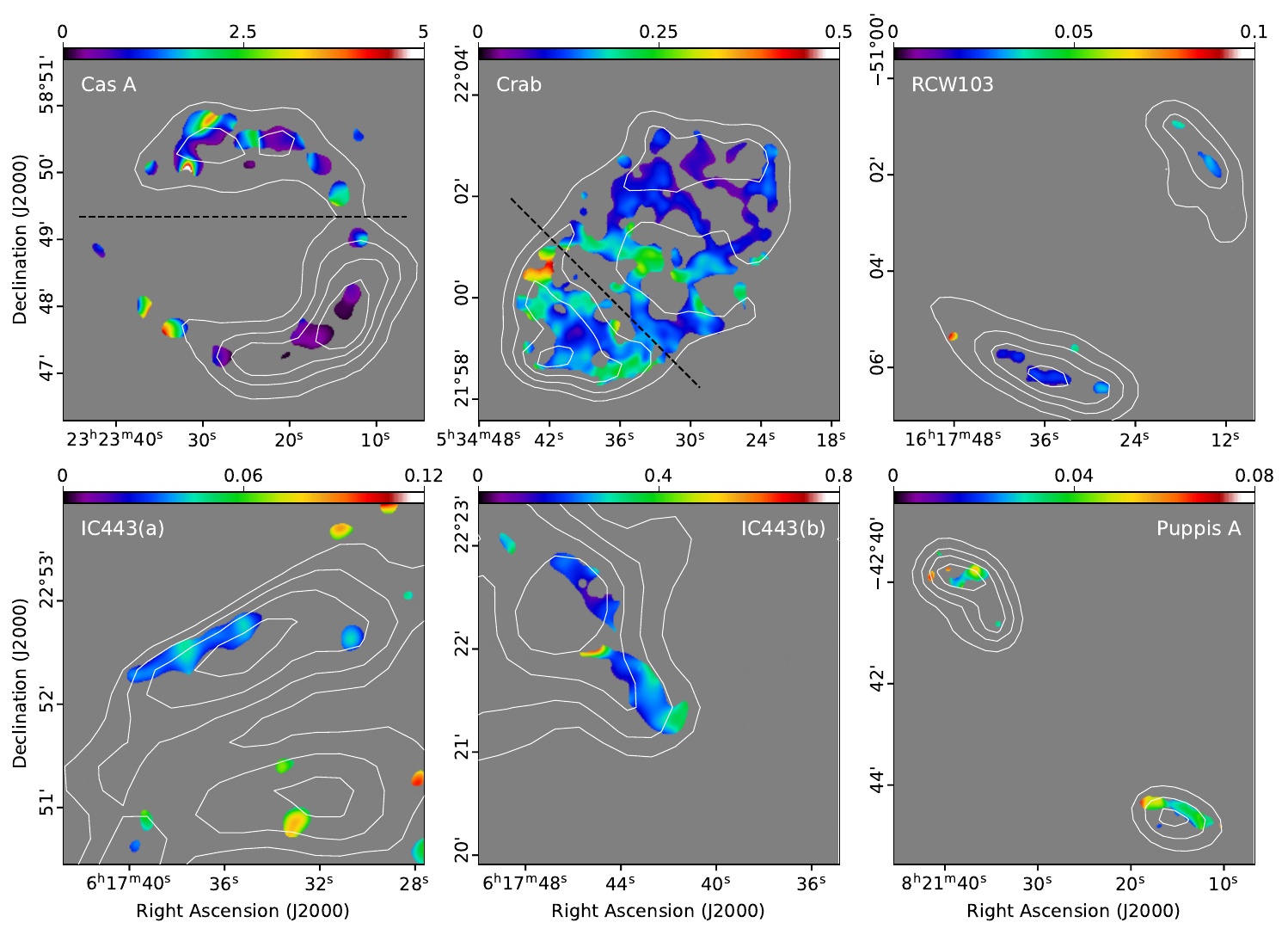}
\end{center}
\caption{Extinction-corrected [P~\emissiontype{II}]/[Fe~\emissiontype{II}] ratio maps for our sample SNRs detected in both the [P~\emissiontype{II}] and [Fe~\emissiontype{II}] emissions. The maps are smoothed by a Gaussian kernel with a sigma of \timeform{5"}. White contours represent the $M_\mathrm{Fe, gas}/M_\mathrm{Fe, dust}$ ratio on a linear scale. Black dashed lines represent the north-south boundary in Cas~A and the northwest-southeast boundary in the Crab Nebula, which are used to classify the data points of these two SNRs in figure~\ref{fig:cor}. {Alt text: Six maps.}} \label{fig:ratio}
\end{figure*}

\begin{table*}[t]
\tbl{Summary of the mid- and far-IR data for our sample SNRs}{
\begin{tabular}{cccccc}
\hline
Name & \multicolumn{4}{c}{Data set} & $T_\mathrm{warm}$\footnotemark[$*$] \\
& WISE & Spitzer & AKARI & Herschel & (K) \\ 
\hline
Cas~A & $22~\micron$ & -- & -- & $70$, $100$, $160$, $250$, $350$, $500~\micron$ & $88{\pm}7$\\
Crab & $22~\micron$ & -- & -- & $70$, $100$, $160$, $250$, $350$, $500~\micron$ & $68{\pm}12$\\
RCW~103 & $22~\micron$ & -- & $90$, $140~\micron$ & $70$, $160$, $250$, $350$, $500~\micron$ & $66{\pm}8$\\
IC~443 & $22~\micron$ & $70~\micron$ & $90$, $140$, $160~\micron$ & -- & $58{\pm}2$\\
Puppis~A & $22~\micron$ & $70~\micron$ & $90$, $140$, $160~\micron$ & -- &  $58{\pm}1$\\
\hline
\end{tabular}} \label{tab:dust}
\begin{tabnote}
\footnotemark[$*$] Warm dust temperatures derived from the SED fits. The temperature for Cas~A is the median of the local SED fits, while those for the other SNRs are derived from the total SED fits (see text for details). \\ 
\end{tabnote}
\end{table*}

\begin{figure*}[t]
\begin{center}
\includegraphics[width=\textwidth]{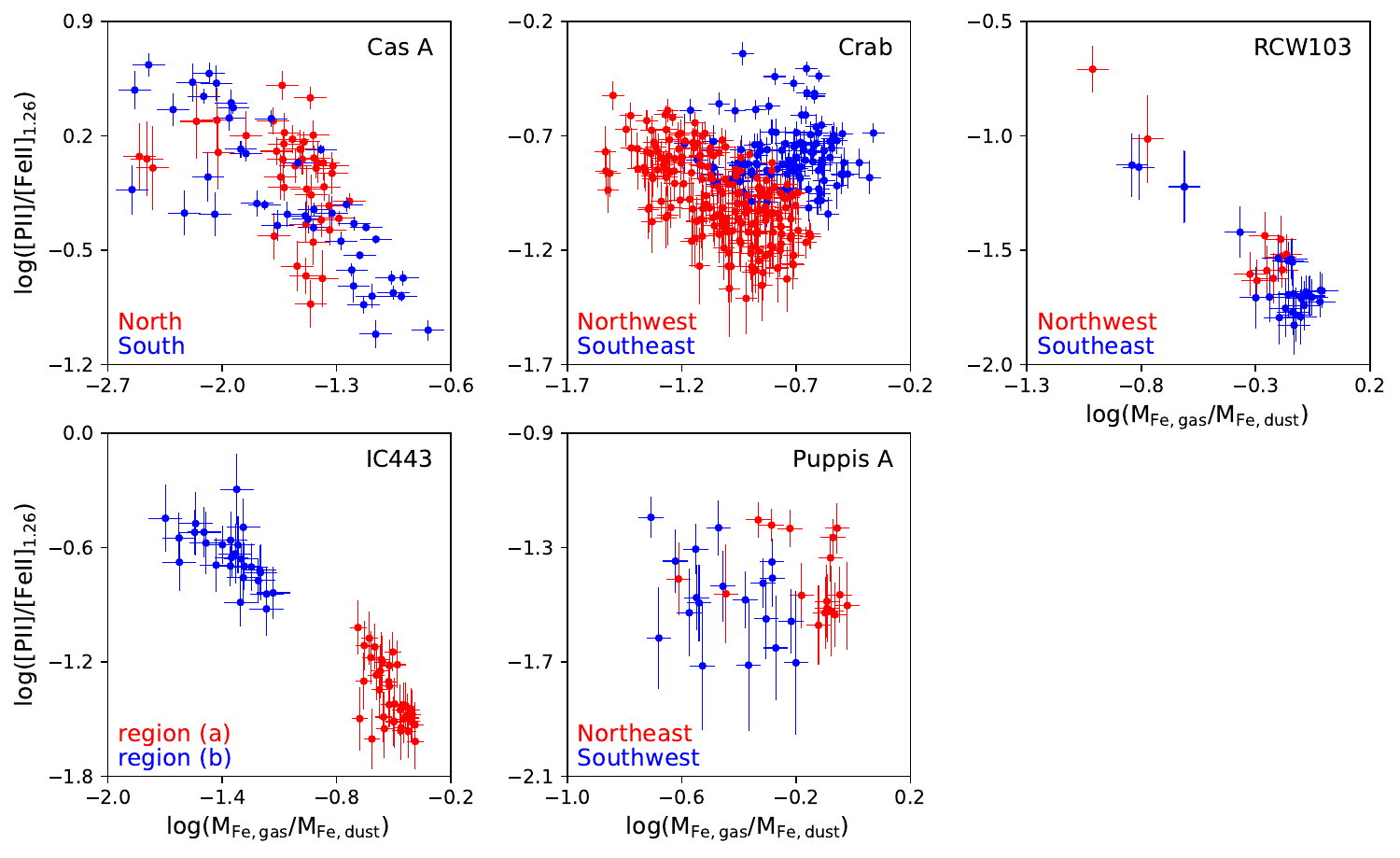}
\end{center}
\caption{Scatter plots between the extinction-corrected [P~\emissiontype{II}]/[Fe~\emissiontype{II}] and $M_\mathrm{Fe, gas}/M_\mathrm{Fe, dust}$ ratios for the SNRs shown in figure~\ref{fig:ratio}. Data points are spatially sampled every $14{\arcsec}$ and color-coded by region in each SNR. The north-south boundary in Cas~A and the northwest-southeast boundary in the Crab Nebula are defined in figure~\ref{fig:ratio}. {Alt text: Five scatter plots.}} \label{fig:cor}
\end{figure*}

\begin{figure*}[t]
\begin{center}
\includegraphics[width=0.8\textwidth]{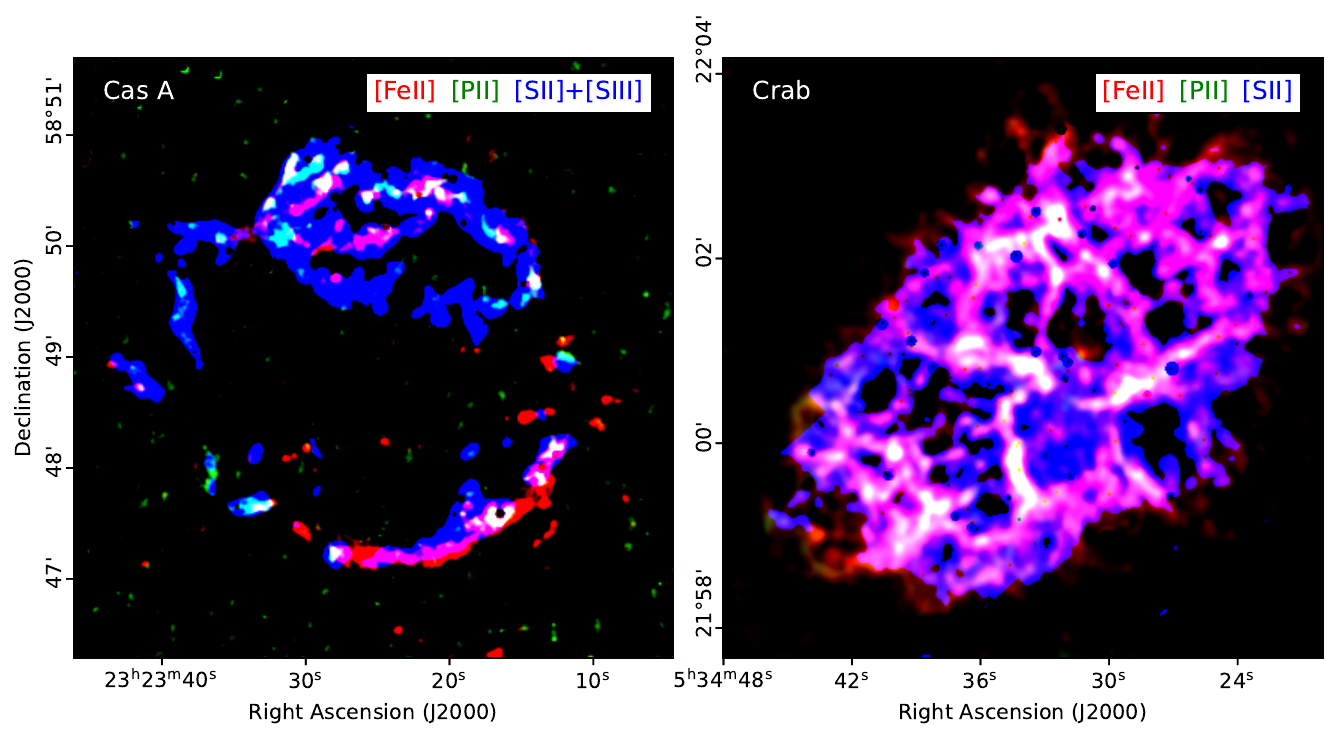}
\end{center}
\caption{Three-color composite images of Cas~A (left) and the Crab Nebula (right). In both panels, red and green show the [Fe~\emissiontype{II}]$_{1.26}$ and [P~\emissiontype{II}] maps, respectively. Blue shows the HST WFC3 F098M (left) and F673N (right) images, dominated by the [S~\emissiontype{III}] $9069$, $9531$~{\AA} and [S~\emissiontype{II}] $1.03~{\micron}$ multiplet emissions and the [S~\emissiontype{II}] $6717$, $6731$~{\AA} emissions, respectively. {Alt text: Two maps.}} \label{fig:rgb}
\end{figure*}

For the SNRs detected in the [P~\emissiontype{II}] and [Fe~\emissiontype{II}]$_{1.26}$ emissions, we investigate spatial variations of the [P~\emissiontype{II}]/[Fe~\emissiontype{II}]$_{1.26}$ ratio as shown in figure~\ref{fig:ratio}. The figure demonstrates that the ratio varies within each SNR, likely caused by asymmetries in ejecta distribution and/or variations in the degree of dust destruction, the latter being due to the fact that P is volatile while Fe is mostly locked in dust grains (e.g.~\cite{sav96}). Although the [Fe~\emissiontype{II}] flux varies significantly with ambient gas density and shock velocity \citep{koo16}, the [P~\emissiontype{II}]/[Fe~\emissiontype{II}]$_{1.26}$ ratio is only weakly affected by these factors, due to the similar excitation conditions of the [P~\emissiontype{II}] and [Fe~\emissiontype{II}]$_{1.26}$ lines as described in section~\ref{sec:abu}. We estimate the degree of dust destruction by evaluating the mass ratio of Fe in the gas phase ($M_\mathrm{Fe, gas}$) to that in the solid phase ($M_\mathrm{Fe, dust}$); since Fe is mostly locked in dust grains in the interstellar and/or circumstellar medium and possibly in supernova ejecta (e.g.,~\cite{sav96}), higher $M_\mathrm{Fe, gas}/M_\mathrm{Fe, dust}$ ratios indicate severer dust destruction by shocks.

We used the mid- and far-IR maps obtained by WISE, Spitzer, AKARI, and Herschel to estimate $M_\mathrm{Fe, dust}$ as summarized in table~\ref{tab:dust}. We resampled all the maps to match a spatial resolution of $14{\arcsec}$, the original pixel size of the Herschel $500~\micron$ map. For the mid-IR map, we adopted the WISE $22~\micron$ band rather than the Spitzer $24~\micron$ band to minimize contamination from the [O~\emissiontype{IV}] $25.89$~$\micron$ and [Fe~\emissiontype{IV}] $25.99$~$\micron$ emissions \citep{del17}. Since dust associated with SNRs is heated by shocks, we separate warm dust from cold dust present in the foreground and/or background by analyzing the IR spectral energy distribution (SED) of each SNR. This assumption may not be directly applicable to the Crab Nebula, where dust heating is primarily due to synchrotron emission. However, in the Crab Nebula, the spatial distribution of the near-IR line emissions more closely matches that of warm dust than of cold dust \citep{gom12, tem24}, suggesting that the warm dust and the gas bright in the [P~\emissiontype{II}] and [Fe~\emissiontype{II}] emissions occupy a similar spatial volume. Hence, the same dust analysis can reasonably be applied to the Crab Nebula.

We fitted local SEDs pixel-by-pixel using a two-temperature modified blackbody model with the emissivity power-law index of two. We also included a power-law component describing synchrotron emission in the model for Cas~A and the Crab Nebula (e.g.,~\cite{del17, del19}). The warm dust temperature is not well constrained for the Crab Nebula due to the dominance of synchrotron and cold dust emissions in the SED \citep{del19}. Therefore, we first performed total SED fits using our model, measuring the flux densities in each IR band within an aperture covering the regions where both the [P~\emissiontype{II}] and [Fe~\emissiontype{II}] emissions are detected. We then performed local SED fits, with the warm dust temperature fixed at the best-fit value in the total SED fits.

For RCW~103, IC~443, and Puppis~A, contamination of the [O~\emissiontype{I}] $63~\micron$ and [C~\emissiontype{II}] $158~\micron$ emissions in the $70~\micron$ and $160~\micron$ bands, respectively, is possible, particularly in shock regions interacting with the interstellar and/or circumstellar medium (e.g.~\cite{bur90, mil21}). To minimize this contamination, as in the case of the Crab Nebula, we first performed total SED fits, and then local SED fits excluding the $70$ and $160~\micron$ bands, where the two dust temperatures are fixed at the best-fit values in the total SED fits. The warm dust temperatures derived from the SED fits are summarized in table~\ref{tab:dust}. We calculate the warm dust mass using the dust mass opacity coefficient given by \citet{dra03}, which is then converted to $M_\mathrm{Fe, dust}$ by assuming an Fe mass fraction of $17\%$, as measured for dust in the interstellar medium \citep{tie05}. The young SNRs Cas~A and the Crab Nebula may contain dust originating from supernova ejecta, and its Fe mass fraction possibly differs from that of the interstellar dust. Indeed, \citet{noz10} predict that an Fe mass fraction of ${\sim}1${\%} for supernova dust in Cas~A, indicating that our estimates of $M_\mathrm{Fe, dust}$ are subject to such systematic uncertainties.

We calculate $M_\mathrm{Fe, gas}$ from our [Fe~\emissiontype{II}]$_{1.26}$ map based on equation~(1) from \citet{koo18} as
\begin{equation}
M_\mathrm{Fe, gas} = 2.13~{\times}~10^{-6}\left(\frac{L_\mathrm{[Fe \emissiontype{II}]1.26}}{L_\odot}\right)\left(\frac{f_\mathrm{Fe \emissiontype{II}}}{1.0}\right)^{-1}\left(\frac{f_{\mathrm{a}^4D_{7/2}, \mathrm{Fe \emissiontype{II}}}}{0.01}\right)^{-1}~M_{\odot},
\label{eq:mfe}
\end{equation}
where $L_\mathrm{[Fe \emissiontype{II}]1.26}$ and $f_{\mathrm{a}^4D_{7/2}, \mathrm{Fe \emissiontype{II}}}$ are the [Fe~\emissiontype{II}]$_{1.26}$ luminosity and the fraction of Fe$^+$ in the $f_{\mathrm{a}^4D_{7/2}}$ level, respectively. Here, we used the Einstein coefficient of $5.27~{\times}~10^{-3}$~s$^{-1}$ derived by \citet{deb11}. We assumed $f_\mathrm{Fe \emissiontype{II}}=1$, a typical value for shock regions bright in near-IR line emission \citep{koo13}, and calculated $f_{\mathrm{a}^4D_{7/2}}$ for each SNR using the mean $n_e$ values listed in table~\ref{tab:abu} and assuming $T_e=10^4$~K.

White contours in figure~\ref{fig:ratio} show the $M_\mathrm{Fe, gas}/M_\mathrm{Fe, dust}$ ratio thus derived. Figure~\ref{fig:cor} shows scatter plots between the [P~\emissiontype{II}]/[Fe~\emissiontype{II}]$_{1.26}$ and $M_\mathrm{Fe, gas}/M_\mathrm{Fe, dust}$ ratios for the five SNRs shown in figure~\ref{fig:ratio}, where the [P~\emissiontype{II}]/[Fe~\emissiontype{II}]$_{1.26}$ maps are spatially sampled every $14\arcsec$ and the data points are color-coded by region within each SNR. Overall, regions with higher $M_\mathrm{Fe, gas}/M_\mathrm{Fe, dust}$ tend to show lower [P~\emissiontype{II}]/[Fe~\emissiontype{II}]$_{1.26}$ ratios, suggesting that dust destruction by shocks may affect the observed line ratios. As the warm dust in Cas~A and the Crab Nebula is primarily attributed to supernova ejecta \citep{del17, del19}, while in the remaining three SNRs it originates from interstellar/circumstellar dust (e.g.~\cite{are10, mil21}), the systematic uncertainties of $M_\mathrm{Fe, dust}$ due to the adopted Fe mass fractions are not expected to significantly affect this trend. In the young SNRs Cas~A and the Crab Nebula, Fe ejecta may reside in the gas phase \citep{noz10, del19}, possibly contributing to the anti-correlation between the [P~\emissiontype{II}]/[Fe~\emissiontype{II}]$_{1.26}$ and $M_\mathrm{Fe, gas}/M_\mathrm{Fe, dust}$ ratios. RCW~103 and IC~443 represent clear anti-correlations in the plot across the observed region. In IC~443, the higher [P~\emissiontype{II}]/[Fe~\emissiontype{II}]$_{1.26}$ ratio observed in region (b) is caused by slow shocks interacting with dense molecular clouds in the southern region (e.g.,~\cite{kok20}), while fast shocks propagating in atomic gases in the northern region (e.g.,~\cite{kok13}) cause severer dust destruction and accordingly lower [P~\emissiontype{II}]/[Fe~\emissiontype{II}]$_{1.26}$ ratios in region (a). In RCW~103, the southeastern region shows systematically lower [P~\emissiontype{II}]/[Fe~\emissiontype{II}]$_{1.26}$ ratios than the northwestern region, suggesting that dust destruction is more efficient in the southeastern region. This is consistent with the presence of hot dust heated by fast shocks in the southwestern region, as inferred from the mid-IR spectra \citep{and11}. Figure~\ref{fig:cor} also shows that the northern, southeastern, and northeastern regions of Cas~A, the Crab Nebula, and Puppis~A, respectively, deviate from the anti-correlation between the [P~\emissiontype{II}]/[Fe~\emissiontype{II}]$_{1.26}$ and $M_\mathrm{Fe, gas}/M_\mathrm{Fe, dust}$ ratios. The origin of this trend in Cas~A and the Crab Nebula will be discussed in the next section. In the case of Puppis~A, since our observations cover only a limited area, large-scale observations of the entire remnant may be needed to investigate the cause of this deviation.



\subsection{Possible origins of the [P~\emissiontype{II}] emission in Cas~A and the Crab Nebula} \label{sec:ori}

Figure~\ref{fig:cor} shows a strong anti-correlation between the [P~\emissiontype{II}]/[Fe~\emissiontype{II}]$_{1.26}$ and $M_\mathrm{Fe, gas}/M_\mathrm{Fe, dust}$ ratios in the southern region of Cas~A and the northwestern region of the Crab Nebula. In addition to dust destruction, Fe-rich ejecta may also contribute to this trend in these two SNRs (e.g.,~\cite{lee17}). On the other hand, no such anti-correlations are found in the northern region of Cas~A and the southeastern region of the Crab Nebula, suggesting that the scatter in the [P~\emissiontype{II}]/[Fe~\emissiontype{II}]$_{1.26}$ ratio may be caused by different mechanisms.

In Cas~A, previous studies reveal that the northern region is relatively rich in S (e.g.,~\cite{mil13}), one of the primary products formed along with P during explosive Ne-burning \citep{rau02, koo13}. The left panel of figure~\ref{fig:rgb} compares the [P~\emissiontype{II}] and [Fe~\emissiontype{II}]$_{1.26}$ maps with the Hubble Space Telescope (HST) WFC3 F098M image tracing the [S~\emissiontype{III}] $9069$, $9531$~{\AA} and [S~\emissiontype{II}] $1.03~{\micron}$ multiplet emissions \citep{fes16}. The figure demonstrates that P and S occupy the same region across the remnant. In particular, in the northern and eastern regions, parts of the shock region are rich in P and S but poor in Fe. This result is consistent with near-IR spectroscopy of Cas~A by \citet{lee17} who find that P- and S-rich ejecta tend to distribute toward the northern and eastern regions. Hence, synthesized P may be preferentially ejected toward these regions together with S, and the observed scatter in the [P~\emissiontype{II}]/[Fe~\emissiontype{II}]$_{1.26}$ ratio in the northern region may reflect variations in the P/Fe abundance ratios of the ejecta.

In the Crab Nebula, deviations from the anti-correlation between the [P~\emissiontype{II}]/[Fe~\emissiontype{II}]$_{1.26}$ and $M_\mathrm{Fe, gas}/M_\mathrm{Fe, dust}$ ratios are evident in the southeastern region. The right panel of figure~\ref{fig:rgb} shows the [P~\emissiontype{II}] and [Fe~\emissiontype{II}]$_{1.26}$ maps, together with the HST WFC3 F673N image tracing the [S~\emissiontype{II}] $6717$, $6731$~{\AA} emissions \citep{lol13}, where we used the F547M image to subtract the continuum emission. The figure reveals that the overall distributions of P, Fe, and S are similar, but a detailed abundance analysis using optical line emissions by \citet{sat12} suggests that the southeastern region represents enhanced S abundance. Combined with our results, this suggests that synthesized P and S may be preferentially ejected into the southeastern region of the Crab Nebula. On the other hand, as previous studies report the detection of the [Ni~\emissiontype{II}] $1.191$~$\micron$ emission in the Crab Nebula \citep{hud90, rud94}, it may also be possible that the observed line flux in the southeastern region is contaminated by the [Ni~\emissiontype{II}] emission. This suggests that the Ni/Fe abundance ratio, rather than the P/Fe abundance ratio, is higher in this part of the remnant. However, these earlier studies did not consider the possibility that the observed line emission originates from [P~\emissiontype{II}] $1.189$~$\micron$, and their spectral resolution was not sufficient to resolve the two lines. High-resolution spectroscopy of the Crab pulsar region by \citet{sol19} reports the detection of the [P~\emissiontype{II}] $1.189$~$\micron$ emission but not the [Ni~\emissiontype{II}] $1.191$~$\micron$ emission. Since these observations were limited to parts of the bright filaments, future spectral mapping with high spectral resolutions is required to investigate these possibilities.

Asymmetric spatial distributions of ejecta in SNRs are not rare, and some mechanisms are proposed to explain this trend, such as hydrodynamical instabilities during the explosion (e.g.,~\cite{lop11}). Figure~\ref{fig:cor} shows that the [P~\emissiontype{II}]/[Fe~\emissiontype{II}]$_{1.26}$ ratio and accordingly $X$(P/Fe) are higher in the northern region of Cas~A compared to the southeastern region of the Crab Nebula by up to an order of magnitude. This difference may be even greater when considering the possible [Ni~\emissiontype{II}] contamination in the [P~\emissiontype{II}] narrow-band image of the Crab Nebula. One possible explanation for this difference is a variation in their progenitor masses; the P/Fe mass ratio for a progenitor mass of $20~M_{\odot}$ is higher than that for $15~M_{\odot}$ by an order of magnitude \citep{rau02}. The progenitor masses of Cas~A and the Crab Nebula are estimated to be $15$--$25$ and ${\sim}9~M_{\odot}$, respectively \citep{you06, owe15}, supporting this scenario. It is also suggested that the progenitor star of Cas~A may have experienced convective-reactive nucleosynthesis between O and C shells, which can produce more odd atomic number elements including P than usual core-collapse supernovae by more than an order of magnitude \citep{rit18, yad20}, and this type of nucleosynthesis is not rare for progenitor masses of $16$--$26~M_{\odot}$ \citep{col18}. Such differences in the progenitor mass may partly contribute to the scatter in the [P~\emissiontype{II}]/[Fe~\emissiontype{II}]$_{1.26}$ ratio in figures~\ref{fig:flux} and \ref{fig:ne}, although we need more sensitive observations to confirm this possibility.

\section{Conclusion} \label{sec:con}
In order to investigate the origin of P in SNRs, we performed large-area [P~\emissiontype{II}] and [Fe~\emissiontype{II}] line mapping of $26$ Galactic SNRs with the IRSF and Kanata telescopes, using the narrow-band filters tuned to the [P~\emissiontype{II}] $1.189~\micron$ and [Fe~\emissiontype{II}] $1.257$ and $1.644~\micron$ lines. There is a possibility that the [Ni~\emissiontype{II}] $1.191~\micron$ line may also contribute to our [P~\emissiontype{II}] narrow-band image, particularly in the case of the Crab Nebula. By combining our data with archival [Fe~\emissiontype{II}] maps from UKIRT, we detected both the [P~\emissiontype{II}] and [Fe~\emissiontype{II}] emissions in five SNRs, only the [Fe~\emissiontype{II}] emission in $15$ SNRs, and no line emissions in the remaining six. Overall, spatial distributions of the [P~\emissiontype{II}] and [Fe~\emissiontype{II}] emissions are similar to each other, both tracing radiative shock regions in our sample SNRs. We estimate $X$(P/Fe) from the [P~\emissiontype{II}]/[Fe~\emissiontype{II}] ratio, to find that $X$(P/Fe) differs significantly from remnant to remnant by up to two orders of magnitude, suggesting that the production rate of P and/or the degree of dust destruction vary among our sample SNRs.

We assess possible effects of dust destruction on the observed [P~\emissiontype{II}]/[Fe~\emissiontype{II}] ratios for the five SNRs detected in both line emissions. We performed IR SED fits for these SNRs using the mid- and far-IR maps obtained by WISE, Spitzer, AKARI, and Herschel to estimate $M_\mathrm{Fe, dust}$. We also derived $M_\mathrm{Fe, gas}$ from our [Fe~\emissiontype{II}] map to estimate the degree of dust destruction using the $M_\mathrm{Fe, gas}/M_\mathrm{Fe, dust}$ ratio. In general, the [P~\emissiontype{II}]/[Fe~\emissiontype{II}] ratio is anti-correlated with the $M_\mathrm{Fe, gas}/M_\mathrm{Fe, dust}$ ratio in the five SNRs, suggesting that the spatial variation of the [P~\emissiontype{II}]/[Fe~\emissiontype{II}] ratio is mainly attributed to dust destruction and possibly to the presence of Fe-rich ejecta. In contrast, the northern and southeastern regions of Cas~A and the Crab Nebula, respectively, show clear deviations from this trend, possibly due to an asymmetric ejection of P during supernova explosions. In the Crab Nebula, enhanced [Ni~\emissiontype{II}] contamination in the southeastern may also contribute to this trend, although future spectral mapping of the remnant is required to distinguish between these scenarios.


\begin{ack}
This research was supported by JSPS KAKENHI Grant Number JP23K22535 and Inamori foundation. The IRSF project was financially supported by the Sumitomo foundation and Grants-in-Aid for Scientific Research on Priority Areas (A) (Nos. 10147207 and 10147214) from the Ministry of Education, Culture, Sports, Science and Technology (MEXT). The operation of IRSF is supported by Joint Development Research of National Astronomical Observatory of Japan, and Optical Near-Infrared Astronomy Inter-University Cooperation Program, funded by the MEXT of Japan. B.-C. K. acknowledges support from the Basic Science Research Program through the NRF of Korea funded by the Ministry of Science, ICT and Future Planning (RS-2023-00277370). R. A. C. was supported by the Institute for Basic Science (IBS-R035-C1).
\end{ack}


\appendix 
\section*{Explosion types and SNR ages of our sample SNRs}

Table~\ref{tab:para} shows the explosion types and SNR ages of our sample SNRs.

\begin{table}[t]
\tbl{Explosion types and SNR ages of our sample SNRs}{
\begin{tabular}{cccc}
\hline
Name & Type\footnotemark[$*$] & Age & References\footnotemark[$\dagger$] \\
& & (yr) & \\
\hline
G1.9$+$0.3 & Ia & 110--170 & (1)\\
Kepler & Ia & 420 & (2)\\
G11.2--0.3 & CC & 2000 & (3)\\
G15.9$+$0.2 & CC & 2900--5700 & (4)\\
Kes~73 & CC & 2000 & (5)\\
Kes~75 & CC & 720 & (6)\\
3C~391 & CC & 19000 & (7)\\
W44 & CC & 20000 & (8)\\
3C~396 & CC & 3000 & (9)\\
3C~397 & Ia & 7000 & (10)\\
G41.5$+$0.4 & CC & -- & (11)\\
W49B & Ia & 5000--6000 & (12)\\
Cas~A & CC & 340 & (13)\\
Tycho & Ia & 450 & (14)\\
Crab & CC & 970 & (15)\\
IC~443 (a) & CC & 20000 & (16)\\
IC~443 (b) & CC & 20000 & (16)\\
Puppis~A & CC & 4000--5300 & (17)\\
G290.1--0.8 & CC & 50000 & (18)\\
G292.0$+$1.8 & CC & 3000--3400 & (19)\\
Kes~17 & CC & 2000--40000 & (20)\\
RCW~86 (a) & Ia & 1800 & (21)\\
RCW~86 (b) & Ia & 1800 & (21)\\
MSH~15--52 & CC & 1600 & (22)\\
RCW~103 & CC & 2000 & (23)\\
G337.2--0.7 & Ia & 750--5000 & (24)\\
G344.7--0.1 & Ia & 6000 & (25)\\
G349.7$+$0.2 & CC & 2800 & (26)\\
\hline
\end{tabular}} \label{tab:para}
\begin{tabnote}
\footnotemark[$*$] CC and Ia indicate core-collapse and Type Ia explosions, respectively. \\
\footnotemark[$\dagger$]
(1) \citet{rey09, rey08},
(2) \citet{rey07, ste02},
(3) \citet{che05, bor16},
(4) \citet{rey06, sas18},
(5) \citet{bor17a},
(6) \citet{got00},
(7) \citet{wil98, che01},
(8) \citet{wol91},
(9) \citet{har99, su11},
(10) \citet{har99, che99},
(11) \citet{kap02},
(12) \citet{zho18},
(13) \citet{kra08a, fes06},
(14) \citet{ste02, kra08b},
(15) \citet{che77, bla17},
(16) \citet{olb01, lee08},
(17) \citet{pet96, may20},
(18) \citet{kas97, kam15},
(19) \citet{cla80, gha05},
(20) \citet{was16, gel13},
(21) \citet{wil11, zha06},
(22) \citet{sew82},
(23) \citet{got97, car97},
(24) \citet{rak06},
(25) \citet{yam12, com10},
(26) \citet{sla02}\\
\end{tabnote}
\end{table}






\end{document}